\documentclass[a4paper,11pt]{article}
\pdfoutput=1 

\usepackage{jcappub} 

\usepackage[T1]{fontenc} 
\usepackage{soul}
\usepackage{multirow}

\title{\boldmath Inert dark matter in three Higgs doublet model: a blind spot narrative}

\author[a]{Amit Dutta Banik,}
\author[b]{Tapoja Jha,}
\author[b,2]{Eija Tanskanen}

\affiliation[a]{Physics and Applied Mathematics Unit, Indian Statistical Institute,\\Kolkata-700108, India}
\affiliation[b]{Sodankyl{\" a} Geophysical Observatory, University of Oulu,\\T{\" a}htel{\" a}ntie 62, Finland}

\emailAdd{amitdbanik@gmail.com}
\emailAdd{tapoja.jha@oulu.fi}
\emailAdd{eija.tanskanen@oulu.fi}

\abstract{We explore the phenomenology of three Higgs doublet scenario, where the scalar potential is augmented by $\mathbb{Z}_{3} \times \mathbb{Z}_{2}$ symmetry making one doublet inert. Thus in effect, our model of interest is two Higgs plus inert Higgs doublet model charged under $\mathbb{Z}_3$ ((2+I)HDM-$\mathbb{Z}_3$) symmetry. We observe a blind spot feature for dark matter direct detection, as the tree-level dark matter-nucleon scattering cross-section vanishes depending on the mass splitting of dark sector particles. We perform a detailed analysis based on vacuum stability, unitarity, relic abundance, and direct detection results on the model. We also perform profile likelihood analysis and constrain the corresponding parameter space.}

\begin{document}
\maketitle
\flushbottom

\section{Introduction}
\label{sec:intro}

Compelling evidences from cosmological and astrophysical observations confirm the existence of dark mater (DM) abundance in the universe \cite{Planck:2018vyg}. The most general and widely accepted candidate for particle DM is known as weakly interacting massive particle or WIMP with annihilation cross-section that satisfy the DM abundance at GeV scale \cite{Jungman:1995df,Bertone:2004pz}. The youngest discovery of Higgs boson~\cite{CMS:2012qbp,ATLAS:2012yve} marks the completion of the standard model (SM) of particle physics.  However, the SM remains incomplete, as it fails to explain origin of neutrino mass, baryon asymmetry in the universe, and not being able to provide a viable DM candidate. This calls for the beyond the standard model (BSM) extensions. While the BSM extension is necessary to account for WIMP like DM candidates, the null detection of dark matter through various direct and indirect search experiments raises concern about the DM interaction with the visible sector. A null detection of DM through direct detection may be realised from cancellation of DM coupling with SM or BSM sector leading to natural blind spot. In the supersymmetric extension of SM, this kind of blind spot like feature can be obtained in case of bino-wino (or bino-Higgsino) mixing \cite{Huang:2014xua,OHare:2015utx,Crivellin:2015bva,Badziak:2015exr,Han:2016qtc}. In non-SUSY extensions of SM, a cancellation in the DM coupling with visible sector can also arise from the model \cite{Cheung:2013dua,Baum:2017enm,Dedes:2014hga}. In the present work, we address a similar conjunction for a BSM extension of the SM with plausible DM candidate as the portal coupling to DM is cancelled out.

\vspace{0.2cm}

\noindent
Two higgs doublet model (THDM) is one of the simplest non-supersymmetric extension of SM with rich phenomenology \cite{Branco:2011iw,Gunion:1989we}. The THDM naturally conserves the oblique electroweak parameter $\rho$ at tree level \cite{Branco:2011iw}. 
The THDM may provide new independent signatures at detection from colliders and can explain BSM phenomenology, still it lacks the potency to provide a feasible DM candidate unless one of the Higgs doublet is considered as inert. The inert higgs doublet is capable of providing a stable WIMP candidate, termed as inert doublet model (IDM) \cite{Deshpande:1977rw,Barbieri:2006dq}. However, the null detection of IDM in direct detection experiments can not be achieved without tuning the Higgs-DM coupling to the value zero. A complete model with null detection of IDM like candidate can be orchestrated considering a three Higgs doublet model (3HDM) where the coupling between DM and SM Higgs can naturally obtain blind spot for direct detection under certain attributes of the model parameters.

\vspace{0.2cm}

\noindent
Simplified extensions of 3HDM with one active and two inert doublets leading to DM are explored extensively in literatures with $\mathbb{Z}_2$ and  $\mathbb{Z}_2\times \mathbb{Z}_2^{\prime}$ symmetry \cite{Keus:2014jha,Keus:2014isa,Keus:2015xya,Cordero-Cid:2016krd,Cordero-Cid:2018man,Cordero-Cid:2020yba,Hernandez-Sanchez:2020aop,Hernandez-Otero:2022dxd,Dey:2023exa}. The scenario with two active and one inert doublets ((2+I)HDM) remains less investigated with the essence of $\mathbb{Z}_2\times \mathbb{Z}_2^{\prime}$ symmetry \cite{Keus:2013hya,Moretti:2015cwa,Merchand:2019bod} and $S_3$ symmetry \cite{Khater:2021wcx,Kuncinas:2023pub}. {The $\mathbb{Z}_3$ symmetric three higgs doublet model is proposed recently} \cite{Bento:2017eti,Chakraborti:2021bpy},  and ``Hermaphrodite DM'' scenario is explored under the $\mathbb{Z}_3$ symmetry assuming one active Higgs doublet \cite{Aranda:2019vda}. Although various independent studies are performed with DM in 3HDM, the null detection phenomena of DM with the realisation of blind spot is not investigated in detail specifically in the context with (2+I)HDM. It is also worth mentioning, a few articles consider some minimal extensions of THDM to achieve null observation of DM direct detection; such as,  Ref.~\cite{Cabrera:2019gaq} considers a fermionic-Higgs portal and shows blind spot is observed without any cancellation among parameters or any fine tuning; one recent article~\cite{Arcadi:2025sxc} considers THDM with pseudoscalar extensions where null observations in direct detection is observed from mutual cancellation of model parameters. However, the present article is the first work in 3HDM scenarios to the best of our knowledge.

\vspace{0.2cm}

\noindent
In the present work, we consider a 3HDM scenario with two active Higgs doublets associated with one inert Higgs doublet charged under $\mathbb{Z}_3\times \mathbb{Z}_2$ symmetry. The $\mathbb{Z}_3$ symmetry between active Higg doublets revokes the possibility of flavour changing neutral current (FCNC) and the $\mathbb{Z}_2$ stabilizes the DM arising from inert Higgs doublet.
The DM candidate in the present model can evade direct detection by canvassing a blind spot region for null detection.
Unlike the bino-wino mixing, the blind spot in the present model is observed when the coupling of DM with the Higgs boson becomes vanishing naturally from the mutual cancellation among model parameters~\cite{Huang:2014xua,OHare:2015utx,Crivellin:2015bva,Badziak:2015exr,Han:2016qtc}.

\vspace{0.2cm}

\noindent
The present article is organised as follows: In section \ref{sec:model}, we provide the detailed framework of the theoretical model. The next section (\ref{sec:bounds}) provides the constraints coming different experimental, theoretical, and observational limits that are applicable in the model. In section \ref{sec:null}, we present the blind spot condition from our scenario. The section \ref{sec:result} gives the final results from theoretical constraints, DM phenomenology and, statistical analysis. Finally, in section \ref{sec:conclusion}, we provide a brief discussion of our findings.

\section{The model}
\label{sec:model}

\noindent We explore a scenario with three Higgs doublets (3HDM) that contains two active and one inert doublets (2+I)HDM-$\mathbb{Z}_3$. The Higgs doublets are charged under discrete $\mathbb{Z}_3$ symmetry, and their transformation under the $\mathbb{Z}_3$ are

\begin{equation}
\label{eq: z3charge}
~~\Phi_{1} \rightarrow \omega \, \Phi_{1}, ~~ \Phi_{2} \rightarrow \omega^{2} \, \Phi_{2}, ~~ \Phi_{3} \rightarrow \Phi_{3}.
\end{equation}

\noindent
An additional $\mathbb{Z}_2$ symmetry is imposed under which only the $\Phi_{3}$ is odd, to ensure the stability as DM candidate, as shown in Table.~\ref{tab1}. Therefore, in the present set-up, the lightest particle of the inert doublet $\Phi_3$ is a viable DM candidate. After spontaneous symmetry breaking (SSB), $\Phi_1$ and $\Phi_2$ attains vacuum expectation value (vev) and the inert doublet remains intact due to unbroken $\mathbb{Z}_2$ symmetry.

 \begin{equation}
~~\Phi_{1} = \begin{pmatrix} \phi_{1}^{+} \\ \frac{v_{1} + h_{1} + i \chi_{1}}{\sqrt{2}} \end{pmatrix}, ~~
\Phi_{2} = \begin{pmatrix} \phi_{2}^{+} \\ \frac{v_{2} + h_{2} + i \chi_{2}}{\sqrt{2}} \end{pmatrix}, ~~
\Phi_{3} = \begin{pmatrix} \chi^{+} \\ \frac{\chi_0 + i \chi_A}{\sqrt{2}} \end{pmatrix}.
\label{fields}
\end{equation}

\begin{table} [ht]
 \centering
   \begin{tabular}{|c|c|c|}
     \hline
     Fields & $\mathbb{Z}_{3}$  &  $\mathbb{Z}_{2}$ \\
     \hline
  $\Phi_{1}$  &  $\omega$       & 1 \\
  $\Phi_{2}$  &  $\omega^{2}$   & 1 \\
  $\Phi_{3}$  &   1     &   $-$1 \\
     \hline
   \end{tabular}
   \caption{Scalar fields and their corresponding $\mathbb{Z}_{3}$ and $\mathbb{Z}_{2}$ charges.}
   \label{tab1}
  \end{table}

\noindent
The complete scalar potential of the three Higgs doublet model (3HDM($v_1,~v_2,~0$)) charged under $SU(2)\times U(1)\times \mathbb{Z}_3 \times \mathbb{Z}_2$ is given as
{\small 
\begin{eqnarray}
V(\Phi_1,~\Phi_2,~\Phi_3) &=& m_{11}^{2}(\Phi_{1}^{\dagger}\Phi_{1}) + m_{22}^{2}(\Phi_{2}^{\dagger}\Phi_{2}) + m_{33}^{2}(\Phi_{3}^{\dagger}\Phi_{3}) + \lambda_{1}(\Phi_{1}^{\dagger}\Phi_{1})^{2} + \lambda_{2}(\Phi_{2}^{\dagger}\Phi_{2})^{2} + \lambda_{3}(\Phi_{3}^{\dagger}\Phi_{3})^{2} 
\nonumber  \\ 
&& + \lambda_{4}(\Phi_{1}^{\dagger}\Phi_{1})(\Phi_{2}^{\dagger}\Phi_{2}) + \lambda_{5}(\Phi_{1}^{\dagger}\Phi_{1})(\Phi_{3}^{\dagger}\Phi_{3}) + \lambda_{6}(\Phi_{2}^{\dagger}\Phi_{2})(\Phi_{3}^{\dagger}\Phi_{3}) + \lambda_{7}(\Phi_{1}^{\dagger}\Phi_{2})(\Phi_{2}^{\dagger}\Phi_{1}) 
\nonumber  \\ 
&& + \lambda_{8}(\Phi_{1}^{\dagger}\Phi_{3})(\Phi_{3}^{\dagger}\Phi_{1})  + \lambda_{9}(\Phi_{2}^{\dagger}\Phi_{3})(\Phi_{3}^{\dagger}\Phi_{2}) + [\lambda_{10}(\Phi_{1}^{\dagger}\Phi_{3})(\Phi_{2}^{\dagger}\Phi_{3}) + \hbox{h.c.}]
\nonumber  \\ 
&& -m_{12}^{2}[(\Phi_{1}^{\dagger}\Phi_{2}) + \hbox{h.c.}].\,
\label{pot}
\end{eqnarray}}

We consider natural flavour conservation scenario (NFC) in Yukawa structure \cite{Glashow:1976nt}. Apart from standard four types NFC Yukawa structure, in 3HDM we have additional $Z$-type democratic NFC \cite{Akeroyd:2016ssd}. In our scenario, after employing additional $\mathbb{Z}_{2}$ symmetry to one Higgs doublet, we have in effect four type NFC Yukawa Lagrangian of THDM, among which we explore type-I and type-II cases. In case of type-I, we assume all fermions couple with $\Phi_{2}$ and none with $\Phi_{1}$, and the corresponding Lagrangian is given by:

\begin{eqnarray}
\mathcal{L}_{y_{1}} &=& -\lambda_{d} \bar{Q}_{L} \Phi_{2} d_{R} - \lambda_{d} \bar{d}_{R} \Phi_{2}^{\dagger} Q_{L}, \\
\mathcal{L}_{y_{2}} &=& -\lambda_{u} \bar{Q}_{L} \Phi_{2}^{c} u_{R} - \lambda_{u} \bar{u}_{R} {\Phi_{2}^{c}}^{\dagger} Q_{L}, \\
\mathcal{L}_{y_{l}} &=& -\lambda_{l} \bar{L}_{l} \Phi_{2} l_{R} - \lambda_{l} \bar{l}_{R} \Phi_{2}^{\dagger} L_{l},
\end{eqnarray}

\noindent
whereas in case of type-II up-type quark couple to $\Phi_{2}$, down-type quark and charged lepton couple to $\Phi_{1}$:

\begin{eqnarray}
\mathcal{L}_{y_{2}} &=& -\lambda_{u} \bar{Q}_{L} \Phi_{2}^{c} u_{R} - \lambda_{u} \bar{u}_{R} {\Phi_{2}^{c}}^{\dagger} Q_{L}, \\
\mathcal{L}_{y_{1}} &=& -\lambda_{d} \bar{Q}_{L} \Phi_{1} d_{R} - \lambda_{d} \bar{d}_{R} \Phi_{1}^{\dagger} Q_{L}, \\
\mathcal{L}_{y_{l}} &=& -\lambda_{l} \bar{L}_{l} \Phi_{1} l_{R} - \lambda_{l} \bar{l}_{R} \Phi_{1}^{\dagger} L_{l}.
\end{eqnarray}

\noindent
The $\mathbb{Z}_{3}$ charge assignment of fermions in case of the above two scenarios are given in Table~\ref{tab2}.

\begin{table}[!htbp]
\centering
\begin{tabular}{|c|cc|}
\hline
\multirow{2}{*}{Fermion} & \multicolumn{1}{c|}{type-I}       & type-II      \\ \cline{2-3} 
                         & \multicolumn{2}{c|}{$\mathbb{Z}_{3}$ charge}     \\ \hline \hline
$Q$                  & \multicolumn{1}{c|}{1}            & 1            \\ \hline
$u$                      & \multicolumn{1}{c|}{$\omega^{2}$} & $\omega^{2}$ \\ \hline
$d$                      & \multicolumn{1}{c|}{$\omega$}     & $\omega^{2}$ \\ \hline
$L$                  & \multicolumn{1}{c|}{$\omega^{2}$} & $\omega$     \\ \hline
$l$                      & \multicolumn{1}{c|}{1}            & 1            \\ \hline
\end{tabular}
\label{tab2}
\caption{$\mathbb{Z}_{3}$ charge assignment featuring NFC in case of first two THDM.}
\end{table}

\noindent
After spontaneous symmetry breaking, the mixing between $\Phi_1$ and $\Phi_2$ leads to five physical scalar particles in the visible sector $h,~H,~H^{\pm},~A$ with physical masses $m_h,~m_H,~m_{H^{\pm}},$ and $m_A$. Physical CP even scalar are obtained after mass diagonalisation of $h_1$ and $h_2$ by an angle $\alpha$. Similarly, CP even scalar and charged scalars are diagnoalised by angle $\beta$. The vevs of the doublets can be rescaled by $v=\sqrt{v_1^2+v_2^2}= 246$ GeV, where $\tan\beta= v_2/v_1$. The inert doublet $\Phi_3$ gives rise to four dark scalars $\chi_0,~\chi^\pm,~\chi_A$. Finally, we realise a scenario where, the scalar sector has fourteen free parameters given as
\begin{equation}
v,~\tan\beta,~\alpha,~m_h,~m_H, ~m_A,~m_{H^\pm},~m_{\chi_0},~m_{\chi_A}, ~m_{\chi^\pm},~\lambda_3,~\lambda_5,~\lambda_6,~\lambda_9.    
\end{equation}
Using the above set of independent parameters of the model, considering $h$ is the SM-like Higgs boson with mass $m_h=125$ GeV, various coupling and terms of the scalar potential can be expressed as follows

\allowdisplaybreaks
\begin{eqnarray}
\label{s1}
\lambda_{1} &=& \frac{m_{h}^{2} \cos^{2}\alpha   + m_{H}^{2} \sin^{2}\alpha  - m_{A}^{2}\sin^{2}\beta  }{2 v^{2} \cos^{2}\beta },\\
\label{s2}
\lambda_{2} &=& \frac{m_{h}^{2}\sin^{2}\alpha   + m_{H}^{2} \cos^{2}\alpha   - m_{A}^{2} \cos^{2}\beta }{2 v^{2} \sin^{2}\beta},\\
\label{s3}
\lambda_{4} &=& \frac{\cos\alpha \sin\alpha (m_{h}^{2} - m_{H}^{2}) - \sin\beta \cos\beta (m_{A}^{2} - 2 m_{H^\pm}^{2})}{v^{2} \cos\beta  \sin\beta},\\
\label{s4}
\lambda_{7} &=& \frac{2(m_{A}^{2} - m_{H^\pm}^{2})}{v^{2}},\\
\label{s5}
\lambda_{8} &=& \frac{(m_{\chi_A}^{2} + m_{\chi_0}^{2} - 2 m_{\chi^\pm}^{2} - v^{2} \lambda_{9}\sin^{2}\beta )}{v^{2} \cos^{2}\beta}, \\
\label{s6}
\lambda_{10} &=& \frac{m_{\chi_0}^{2} - m_{\chi_A}^{2}}{v^{2} \sin 2\beta}, \\
\label{s7}
m_{12}^{2} &=& m_{A}^{2}{\cos\beta \sin\beta},\\
\label{s8}
m_{33}^{2} &=& \frac{(2 m_{\chi^\pm}^{2} - 
v^{2} \lambda_{5}\cos^{2}\beta  - v^{2} \lambda_{6}\sin^{2}\beta )}{2}.
\end{eqnarray}

\section{Theoretical and experimental constraints}
\label{sec:bounds}

We begin our analysis by employing different theoretical and experimental constraints on the model parameters. We examine theoretical constraints coming from perturbativity, vacuum stability and, unitarity; whereas, for the experimental ones we make sure the bounds of BSM masses maintained from electroweak precision observables (EWPO), bounds on masses of charged particles from LEP, and flavour constraints from, e.g., ${\rm Br}(B \rightarrow X_{s}\gamma)$ etc.

\subsection*{Perturbativity}

\noindent
To ensure the theory remains calculable perturbatively, all the quartic couplings of the  scalar potential in Eq.~(\ref{pot}) should follow the condition $\lambda_{i} \leq |4\pi|$.

\subsection*{Stability of potential}

\noindent
For the stability condition, the scalar potential should be bounded from below in all directions, i.e., 2, 3, and, 4-field directions in the field space. We have calculated the copositivity criteria~\cite{Kannike:2012pe} to obtain the conditions and have checked that bounds coming from 2-field directions are the most stringent~\cite{Chakrabortty:2013mha}. The conditions are given by the following set of equations:

\allowdisplaybreaks
\begin{eqnarray}
\lambda_{1,2,3} & \geq & 0, \\ 
\lambda_{4} + 2 \sqrt{\lambda_{1} \lambda_{2}} & \geq & 0,\\
\lambda_{5} + 2 \sqrt{\lambda_{1}\lambda_{3}} & \geq & 0,\\
\lambda_{6} + 2 \sqrt{\lambda_{2}\lambda_{3}} & \geq & 0,\\
\lambda_{4} + \lambda_{7} + 2 \sqrt{\lambda_{1} \lambda_{2}} & \geq & 0,\\
\lambda_{5} + \lambda_{8} + 2 \sqrt{\lambda_{1}\lambda_{3}} & \geq & 0,\\
\lambda_{6} + \lambda_{9} + 2 \sqrt{\lambda_{2}\lambda_{3}} & \geq & 0.
\end{eqnarray}

\subsection*{Unitarity constraints}

\noindent
To remain predictive at high energy scale, the potential must obey the following unitarity conditions:

\allowdisplaybreaks
\begin{flalign}
&\lambda_{1} \leq 4\pi,&\\
&\lambda_{2} \leq 4\pi,&\\
&\lambda_{4} \leq 8\pi,&\\
&\lambda_{5} \pm \lambda_{8} \leq 8\pi,&\\
&\lambda_{6} \pm \lambda_{9} \leq 8\pi,&\\
&\lambda_{4} + 2\lambda_{7} \leq 8\pi,&\\
&\lambda_{4} - \lambda_{7} \leq 8\pi,&\\
&\lambda_{5} + \lambda_{6} \pm \sqrt{4\lambda_{10}^{2} + (\lambda_{5} - \lambda_{6})^{2}} \leq 16\pi,&\\
&2\lambda_{3} + \lambda_{4} + \lambda_{7} \pm \sqrt{8\lambda_{10}^{2} + (- 2\lambda_{3} + \lambda_{4} + \lambda_{7})^{2}} \leq 16\pi,&\\
&\lambda_{5} + \lambda_{6} + 2(\lambda_{8} + \lambda_{9}) \pm \sqrt{36\lambda_{10}^{2} + (\lambda_{5} - \lambda_{6} + 2\lambda_{8} - 2\lambda_{9})^{2}} \leq 16\pi,&\\
&{\rm Root}(x^{3} - 2(\lambda_{1} + \lambda_{2} + \lambda_{3})x^{2} - \{\lambda_{7}^{2} + \lambda_{8}^{2} + \lambda_{9}^{2} -4(\lambda_{1}\lambda_{2} + \lambda_{2}\lambda_{3} + \lambda_{1}\lambda_{3})\}x \nonumber \\&+ 2(\lambda_{1}\lambda_{9}^{2} + \lambda_{2}\lambda_{8}^{2} + \lambda_{3}\lambda_{7}^{2} - \lambda_{7}\lambda_{8}\lambda_{9} - 4\lambda_{1}\lambda_{2}\lambda_{3}))\leq 8\pi,&\\
&{\rm Root}(x^{3} - 6(\lambda_{1} + \lambda_{2} + \lambda_{3})x^{2} + \{-(\lambda_{7}^{2} + \lambda_{8}^{2} + \lambda_{9}^{2}) - 4(\lambda_{4}^{2} + \lambda_{5}^{2} + \lambda_{6}^{2} + \lambda_{4}\lambda_{7} + \lambda_{5}\lambda_{8} + \lambda_{6}\lambda_{9}) \nonumber \\&+ 36(\lambda_{1}\lambda_{2} + \lambda_{2}\lambda_{3} + \lambda_{1}\lambda_{3})\}x + 6\{\lambda_{1}(\lambda_{9} + 2\lambda_{6})^{2}) + \lambda_{2}(\lambda_{8} + 2\lambda_{5})^{2}) + \lambda_{3}(\lambda_{7} + 2\lambda_{4})^{2}\} \nonumber \\&-2(\lambda_{7} + 2\lambda_{4})(\lambda_{8} + 2\lambda_{5})(\lambda_{9} + 2\lambda_{6}) - 216\lambda_{1}\lambda_{2}\lambda_{3} \leq 8\pi.&
\end{flalign}

\subsection*{SM-like Higgs}

\noindent
In our entire analysis, we maintain Higgs alignment limit, i.e., $\beta - \alpha = 0$~\cite{Das:2019yad}, where $h$ is the SM Higgs boson with physical mass 125 GeV.

\subsection*{Electoweak (EW) precision constraints}

\noindent
The Peskin-Takeuchi $S$, $T$, $U$ parameters~\cite{Peskin:1991sw, Grimus:2008nb, Grimus:2007if, Bertolini:1985ia} are crucial for the validity of any new physics with extended scalar sector. The precision test of EW observables. e.g., gauge boson mass, couplings can be performed in one loop correction of two-point functions of guage bosons. The recent constraints on these parameters~\cite{ParticleDataGroup:2022pth, Batra:2025amk, Coleppa:2025qst} with $\Delta U =0$ is given as

\begin{equation} 
\Delta S = -0.01 \pm 0.07, \, \Delta T = 0.04 \pm 0.06, 
\end{equation}

\noindent
where, $\Delta X = \Delta X^{\rm THDM} + \Delta X^{\rm IDM};\, X = S, T$~\cite{Merchand:2019bod, Hmissou:2025uep}. Detailed analytical expressions can be found in Ref.~\cite{Moretti:2015cwa}. 
 In our work, using the limits from stability conditions discussed in section~\ref{subsec:theory_const}, we have considered 1 GeV mass difference among visible scalars. For the dark sector particles, we consider small mass splitting 5 GeV. Therefore, constraints coming from EWPO is safely evaded.

\subsection*{Constraints on scalar from LEP}

\noindent
The width of the SM $Z$ and $W$ bosons are being precisely measured at LEP~\cite{TevatronElectroweakWorkingGroup:2010mao, Boline:2011qf, ALEPH:2013dgf}. In our analysis, the BSM masses being heavy ($\geq$ 500 GeV) the on-shell decay of $Z$, $W$ to any BSM scalars are prohibited from LEP data. Furthermore, we have also taken into account the constraints on dark masses and their corresponding differences from LEP-II MSSM~\cite{Lundstrom:2008ai}. The set of conditions shows the following mass limits are excluded:

\begin{equation}
 m_{\chi_0} < 80 \,{\rm GeV}, \, m_{\chi_A} < 100 \,{\rm GeV},\, m_{\chi_A} - m_{\chi_0} > 8\, {\rm GeV}.
\end{equation}

\noindent
We have taken DM mass starting from 500 GeV and the corresponding difference is 5 GeV, so the above constraints are well maintained. From Ref.~\cite{ALEPH:2013htx}, bounds on the mass of charged-Higgs above 80 GeV have been well maintained in our analysis.

\subsection*{Flavour constraints}

\noindent
Among all the flavour observables, ${\rm Br}(B \rightarrow X_{s}\gamma)$ gives the most stringent constraint on the mass of $H^{\pm}$; the experimental value is (3.32 $\pm$ 0.15) $\times 10^{-4}$, whereas the corresponding SM value is (3.34 $\pm$ 0.22) $\times 10^{-4}$~\cite{HFLAV:2016hnz}. Since the dark $\chi^{\pm}$ does not couple to fermions, the flavour observables are insulated from the contributions of dark charged particle. Therefore, the bounds from THDM analysis will be applicable here. Implementing the deviation the Ref.~\cite{Misiak:2017bgg} shows for type-II, at 95\% C.L. the lower bound on $m_{H^{\pm}}$ is 570-800 GeV. Another article around same time~\cite{Arbey:2017gmh} shows slightly less stringent bound on $m_{H^{\pm}}$, which is for any $\tan\beta$, $m_{H^{\pm}} \geq$ 600 GeV, whereas for $\tan\beta$ < 1, the bound is slightly more constraint, $m_{H^{\pm}} \geq$ 650 GeV. In this article, we have considered, $m_{H^{\pm}} \sim$ 1 TeV, and thus evading the aforementioned constraint.




\subsection*{DM relic density and direct detection}

\noindent
We consider $\chi_0$ as the lightest particle in our analysis. The DM relic abundance obtained from PLANCK is $\Omega_{\rm DM}h^2=0.1198 \pm 0.0012$~\cite{Planck:2018vyg}. We implement the model file generated from \texttt{FeynRules}~\cite{Alloul:2013bka} in \texttt{micrOMEGAs}~\cite{Belanger:2013oya} to determine the abundance of the DM candidate. Recent experimental data of LZ collaborations~\cite{LZ:2024zvo} provides more stringent constraint on spin-independent (SI) cross section as function of WIMP mass. At 90\% C.L., the strongest SI exclusion is given by 2.2 $\times 10^{-48}~{\rm cm}^{2}$. We evaluate the SI of DM scattering off the nucleon using \texttt{micrOMEGAs}.

\section{Blind spot for dark matter direct detection}
\label{sec:null}

The DM-nucleon elastic  SI scattering cross-section in the present model is given as
\begin{eqnarray} \label{cs_el_2hdm}
\sigma_{\rm SI}^{} \,\,\simeq\,\,
\frac{m_N^2\,}{\pi\bigl(m_{\chi_0}+m_N^{}\bigr)^2} \Biggl(\frac{\lambda_h^{}\,g_{NNh}^{}}{m_h^2}
+ \frac{\lambda_H^{}\,g_{NNH}^{}}{m_H^2}\Biggr)^{\!2} \,\,.
\end{eqnarray}
Here, $m_{N}$ is the nucleon mass and 
DM coupling with Higgs $h$ and heavy scalar $H$ are denoted by $\lambda_{h}$ and $\lambda_{H}$. In a similar fashion, $g_{NNh}$ ($g_{NNH}$)
denotes effective nucleon Higgs coupling \cite{He:2008qm} 
\begin{eqnarray} \label{g_nnh}
g_{NNh}^{} \,\,\simeq\,\, \bigl(1.217\,k_d^{h}+0.493\, k_u^{h}\bigr)\times10^{-3} \,\, \nonumber \\
g_{NNH}^{} \,\,\simeq\,\, \bigl(1.217\,k_d^{H}+0.493\, k_u^{H}\bigr)\times10^{-3} \,\,
\end{eqnarray}
where the factors $k_u^{x},~k_d^{x};~x=h,~H$
are determined by the choice of 
the Yukawa structures. 
For example,
in the type-I THDM model these factors are 
\begin{eqnarray}  \label{factors1}
k_u^h  \,\,=k_d^h \,\,=\,\, \frac{\sin\alpha}{\sin\beta} \,\,, \hspace{2em}
k_u^H  \,\,=k_d^H \,\,=\,\, \frac{\cos\alpha}{\sin\beta} \,\,.
\end{eqnarray}
Similarly, for type-II THDM we find  
\begin{eqnarray}  \label{factors2}
k_u^h  \,\,=\,\, \frac{\sin\alpha}{\sin\beta} \,\,, \hspace{2em}
k_d^h \,\,=\,\,  \frac{\cos\alpha}{\cos\beta}
\,\,, \hspace{2em}
k_u^H  \,\,=\,\, \frac{\cos\alpha}{\sin\beta} \,\,, \hspace{2em}
k_d^H \,\,=\,\, -\frac{\sin\alpha}{\cos\beta}
\,\,.
\end{eqnarray}
In the present formalism, the contribution from beyond SM scalar $H$ in the expression of $\sigma_{\rm SI}$ can be safely ignored for large value of $m_H$, and the reduced scattering cross-section is
\begin{eqnarray} \label{cs_el_hdm}
\sigma_{\rm SI}^{h} \,\,\simeq\,\,
\frac{m_N^2\,}{\pi\bigl(m_{\chi_0}+m_N^{}\bigr)^2} \frac{\lambda_h^{2}\,g_{NNh}^{2}}{m_h^4} \,\,.
\label{crssec}
\end{eqnarray}
The DM-Higgs coupling $\lambda_h$ in the present model is expressed as
\begin{eqnarray}
\lambda_h=\frac{1}{v^2}\bigg[(\lambda_5\cos^2\beta+\lambda_6\sin^2\beta)\frac{v^2}{2} -(m^2_{\chi^{\pm}}-m^2_{\chi_0})\bigg]\, .
\label{coup1}
\end{eqnarray}
In the above expression, the dependence of parameter $\beta$ can be removed
with the assumption $\lambda_5=\lambda_6$. Therefore, we consider only the model parameter space that satisfies $\lambda_5=\lambda_6$ for simplicity. Thus the effective condition for the blind spot of dark matter direct detection in our model is $\lambda_h=0$ or
\begin{eqnarray}
\lambda_5\frac{v^2}{2}=m^2_{\chi^{\pm}}-m^2_{\chi_0}\, .
\label{blind}
\end{eqnarray}

\section{Results}
\label{sec:result}
In this section, we present the results that satisfy different constraints mentioned in sections~\ref{sec:bounds} and~\ref{sec:null}.

\subsection{Theoretical constraint}
\label{subsec:theory_const}

\noindent
Before we discuss the DM phenomenology of (2+I)HDM-$\mathbb{Z}_3$, we first constrain the model parameters from theoretical limits. Following are the ranges of our initial parameter space scanning:

\allowdisplaybreaks
\begin{flalign}
& \hspace*{4cm}1\leq \tan\beta \leq 25;~\beta=\alpha,& \nonumber \\ 
&\hspace*{4cm}\lambda_{3}=0.1 \, ,& \nonumber\\
&\hspace*{4cm}-2 \leq \lambda_{5} (=\lambda_{6})\leq2 \, ,&\nonumber\\
&\hspace*{4cm}-2 \leq \lambda_{9}\leq2 \, ,&\nonumber\\
&\hspace*{4cm}m_{H}=1000~{\rm GeV}\, ,&\nonumber\\
&\hspace*{4cm}-200~{\rm GeV} \leq \Delta M (=m_{A}-m_{H})\leq 50~{\rm GeV}; m_{A}=m_{H^\pm}\, ,&\nonumber\\
&\hspace*{4cm}500~{\rm GeV}\leq m_{\chi_0} \leq 1500~{\rm GeV}\, ,&\nonumber\\
&\hspace*{4cm}m_{\chi_A}=m_{\chi^\pm}=m_{\chi_0}+5~{\rm GeV}.&
\label{param}
\end{flalign}

\begin{figure}[h!tbp]
    \begin{center}
    \hspace*{-2cm}
        \includegraphics[width=0.5\textwidth]{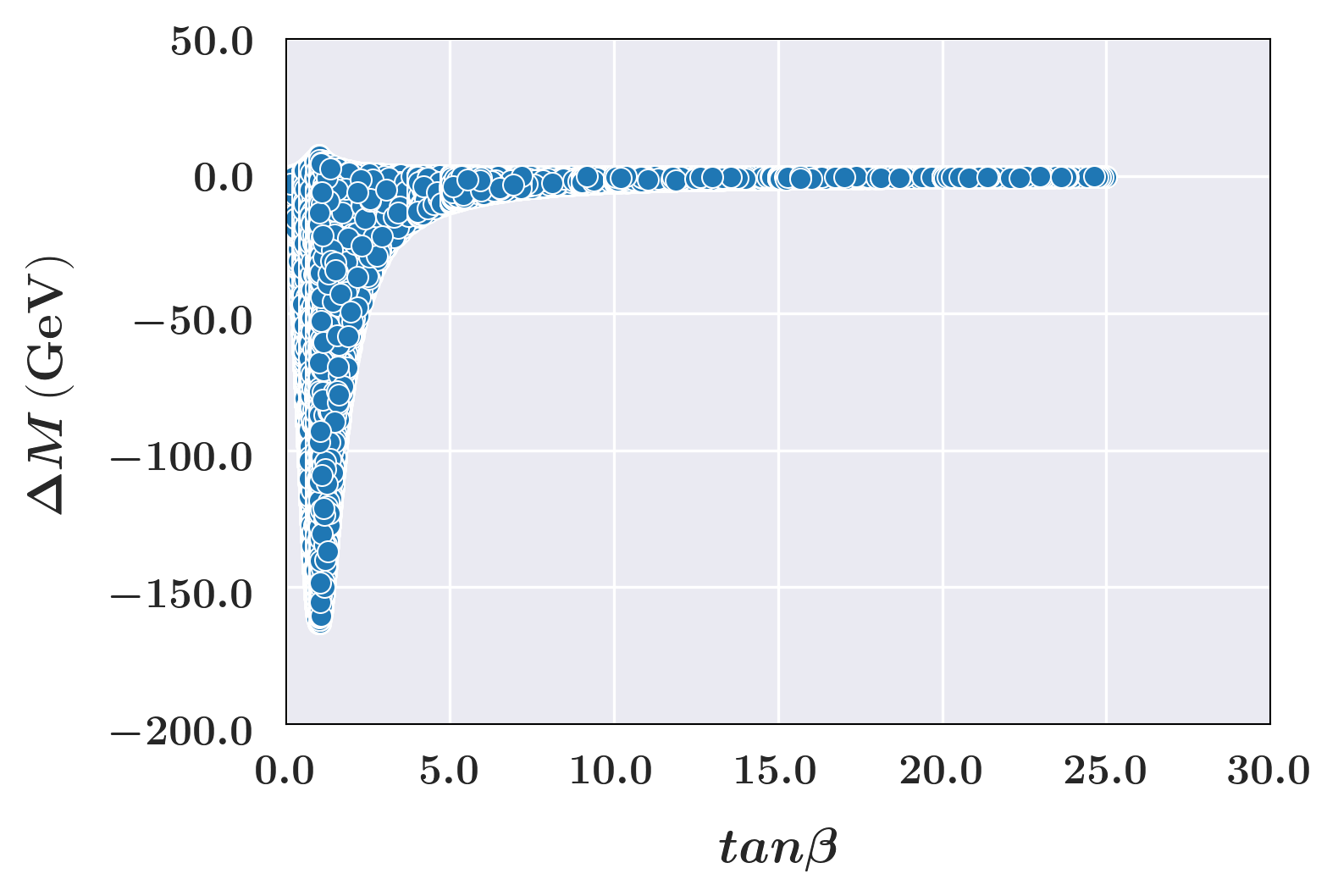}
        \includegraphics[width=0.5\textwidth]{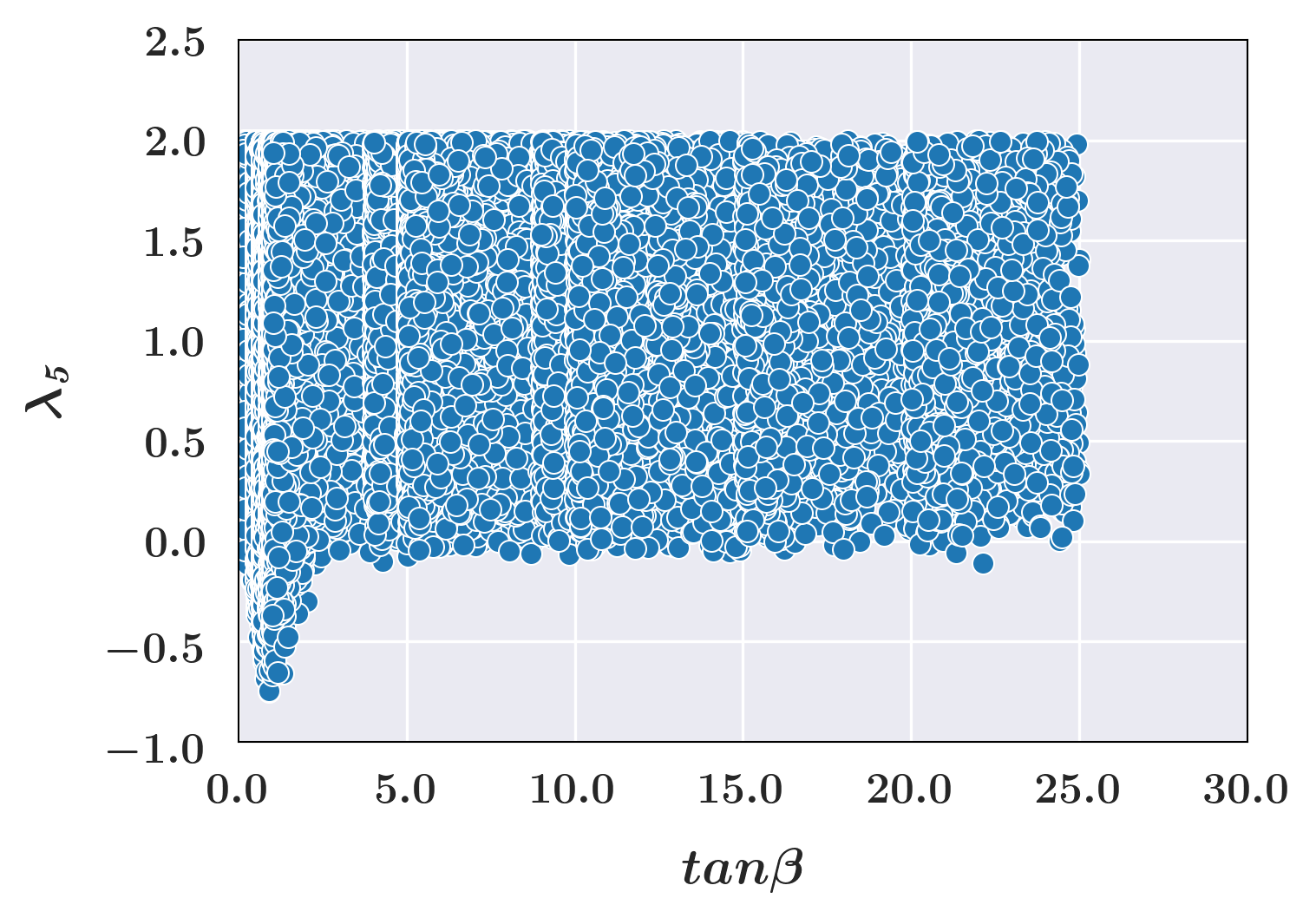}
  \hspace*{-2cm}\includegraphics[width=0.5\textwidth]{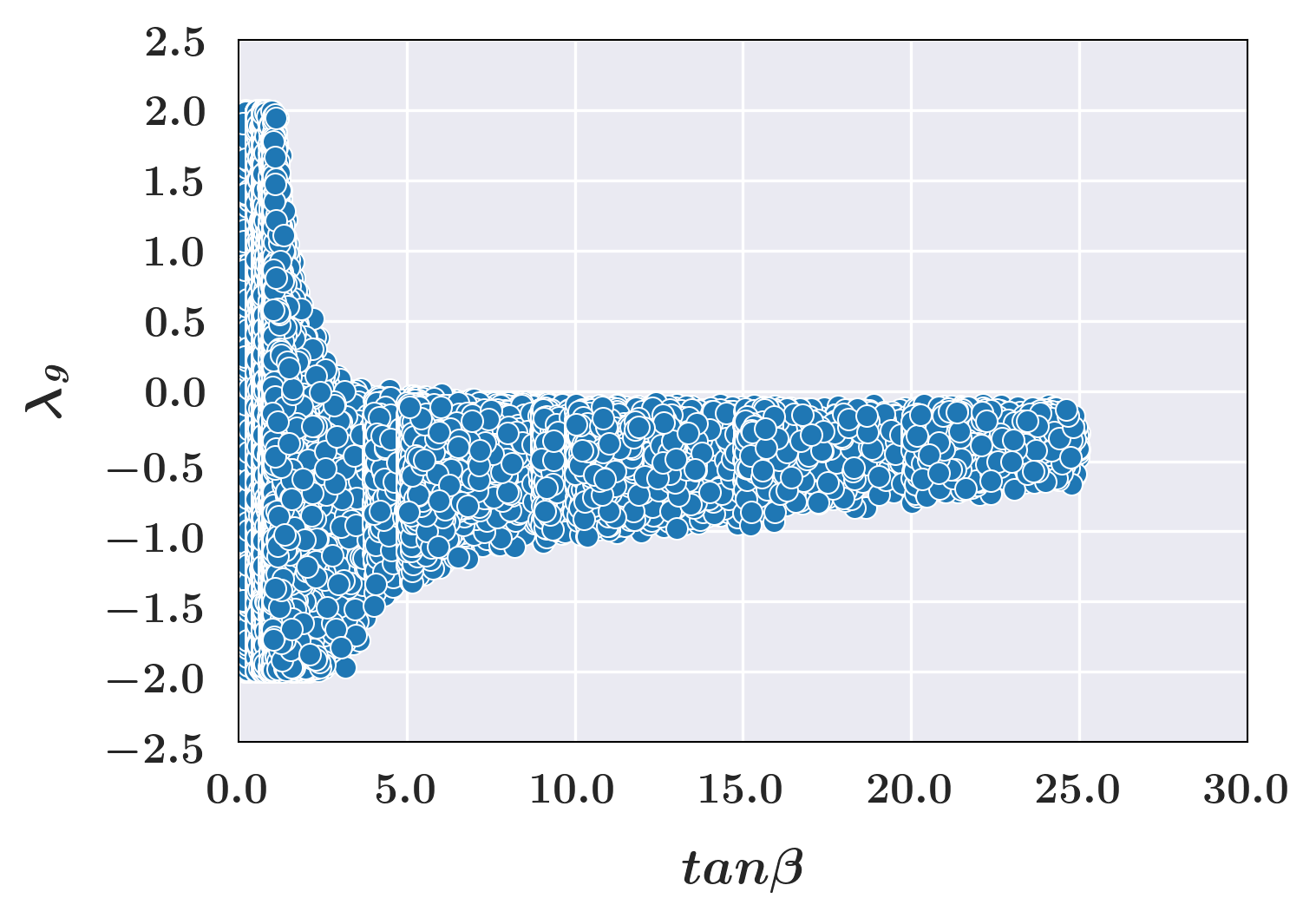}
        \includegraphics[width=0.53\textwidth]{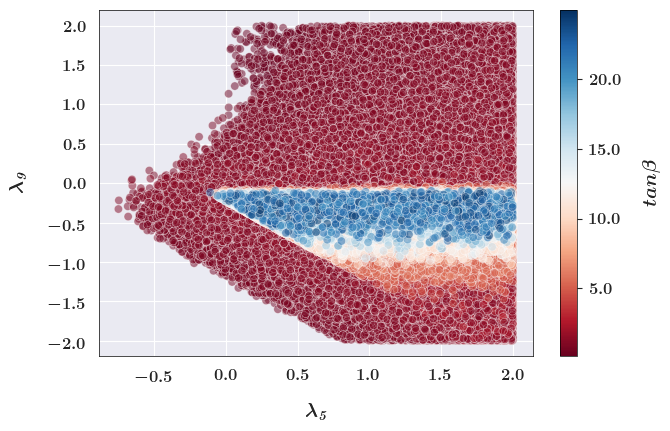}
        \includegraphics[width=0.60\textwidth]{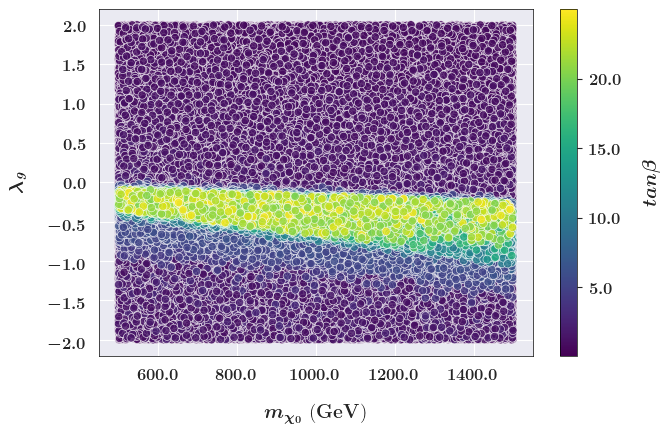}
            \caption{Plots presenting different results obtained from bound from below and unitarity condition of the scalar potential. In row 1, the left panel shows the variation of mass difference of BSM CP-even Higgs and BSM charged (or, pseudoscalar) in GeV as function of $\tan\beta$; the right panel shows the variation of $\lambda_{5}$ with $\tan\beta$. In the next row, the left panel presents the nature of variation of $\lambda_{9}$ as function of $\tan\beta$; the right one represents the allowed region on $\lambda_{5} - \lambda_{9}$ plane in terms of $\tan\beta$. The last one depicts the allowed region in $\lambda_{9}-m_{\chi_0}$ GeV, as function of $\tan\beta$.}
        \label{fig1}
    \end{center}
\end{figure}

\noindent
In Fig.~\ref{fig1}, we obtain the available model parameter space based on stability and unitary constraints considering Eq.~(\ref{param}). The scalar sector equations Eqs.~(\ref{s1})-(\ref{s8})
accompanied by the stability and unitarity limits drives the mass splitting $\Delta M$ to zero for larger $\tan\beta$ values as depicted in Fig.~\ref{fig1}. The dark sector coupling $\lambda_5$ is found to be positive for most of the region and becomes negative for small $\tan\beta$. This follows from the relation, in Eq.~(\ref{s5}), $\lambda_{8} = (m_{\chi_A}^{2} + m_{\chi_0}^{2} - 2 m_{\chi^\pm}^{2} - v^{2} \lambda_{9}\sin^{2}\beta  )/(v^{2} \cos^{2}\beta)$ when simplified to $\lambda_8+\tan^2\beta\lambda_9=\lambda_5/(2\cos^2\beta)$. Therefore, for $\lambda_9<0$ and small $\tan\beta$, $\lambda_5<0$ is achieved for a small region of parameters. 
The variation of $\lambda_9$ coupling with $\tan\beta$ is therefore justified, requiring smaller $\lambda_9$ values with increased $\tan\beta$ and vice-versa. this feature is also shown in the $\lambda_5-\lambda_9$ plane showing larger $\tan\beta$ restricting the dark sector parameters significantly when constraints from stability and unitarity are imposed. The same Eq.~(\ref{s5}) can also be expressed as $\lambda_{8} = \{(m_{\chi_0}^{2} - m_{\chi^\pm}^{2}) - v^{2} \lambda_{9}\sin^{2}\beta \}/(v^{2} \cos^{2}\beta)$, or $\lambda_{8}  = - (1 + \tan^{2}\beta)\,(m_{\chi^{\pm}}^{2} - m_{\chi_0}^{2})/v^{2} - \lambda_{9}\tan^{2}\beta$, so for large $\tan\beta$ if $\lambda_9$ is positive $\lambda_8$ will be largely negative breaking the perturbative limit. Therefore, all the expressions obtained from theoretical constraint push the bound on $\lambda_{8}$ to be less largely negative; thus pushing the value of $\lambda_{9}$ to be more negative within the perturbative limit. So as the DM mass increases, the difference $(m_{\chi^{\pm}}^{2} - m_{\chi_0}^{2})$ increases, thus $\lambda_{9}$ becomes more negative for higher DM mass value to prevent $\lambda_8$ to be largely negative. The last plot of Fig.~\ref{fig1} represents this dependence among the shown parameters. The plots shown in Fig.~\ref{fig1} depends only on scalar sector and applicable for both type-I and type-II models considered.

\subsection{DM phenomenology}
\label{subsec:DMpheno}

\begin{figure}
    \begin{center}
        \includegraphics[width=0.47\textwidth]{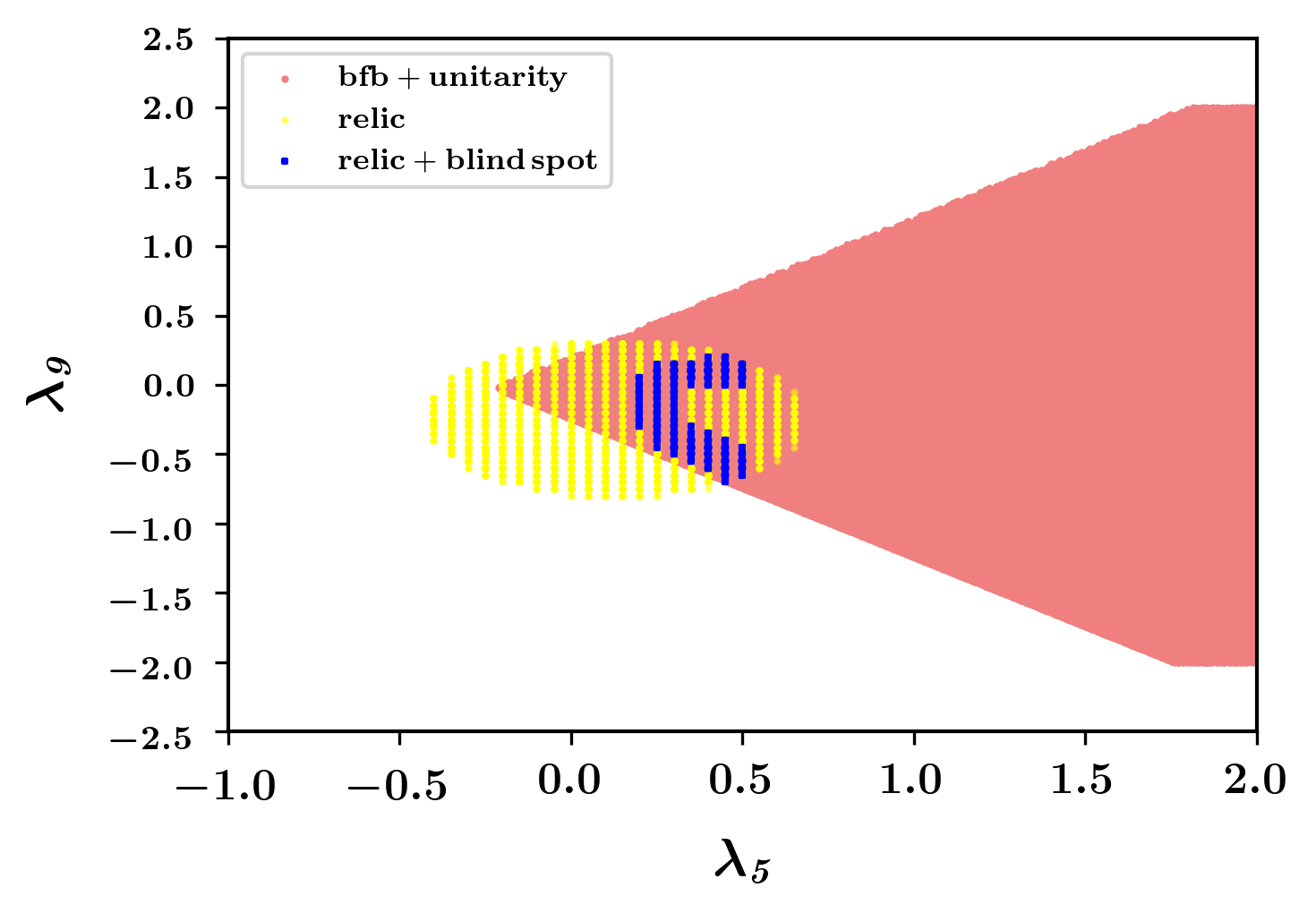}
        \includegraphics[width=0.50\textwidth]{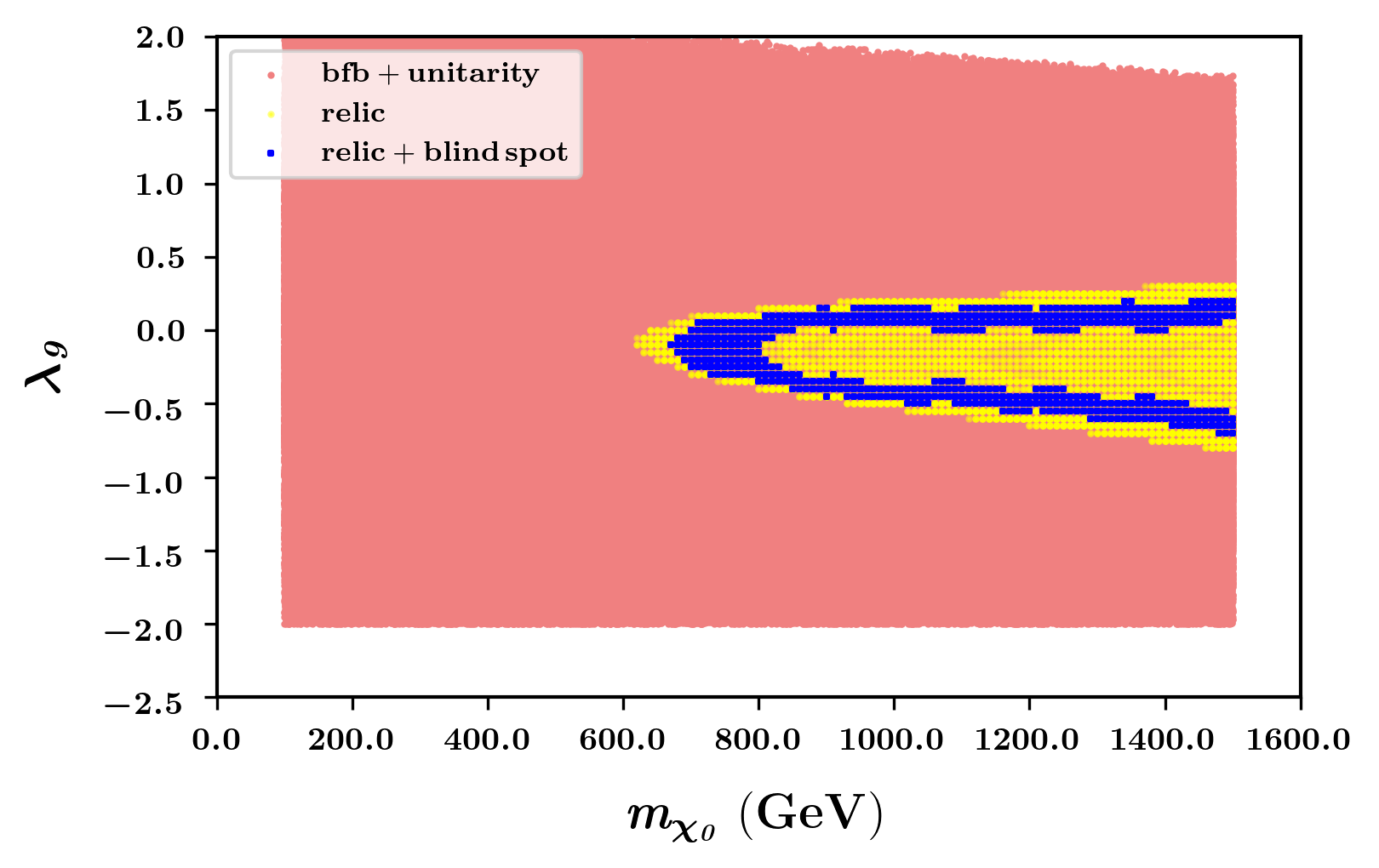}
        \includegraphics[width=0.47\textwidth]{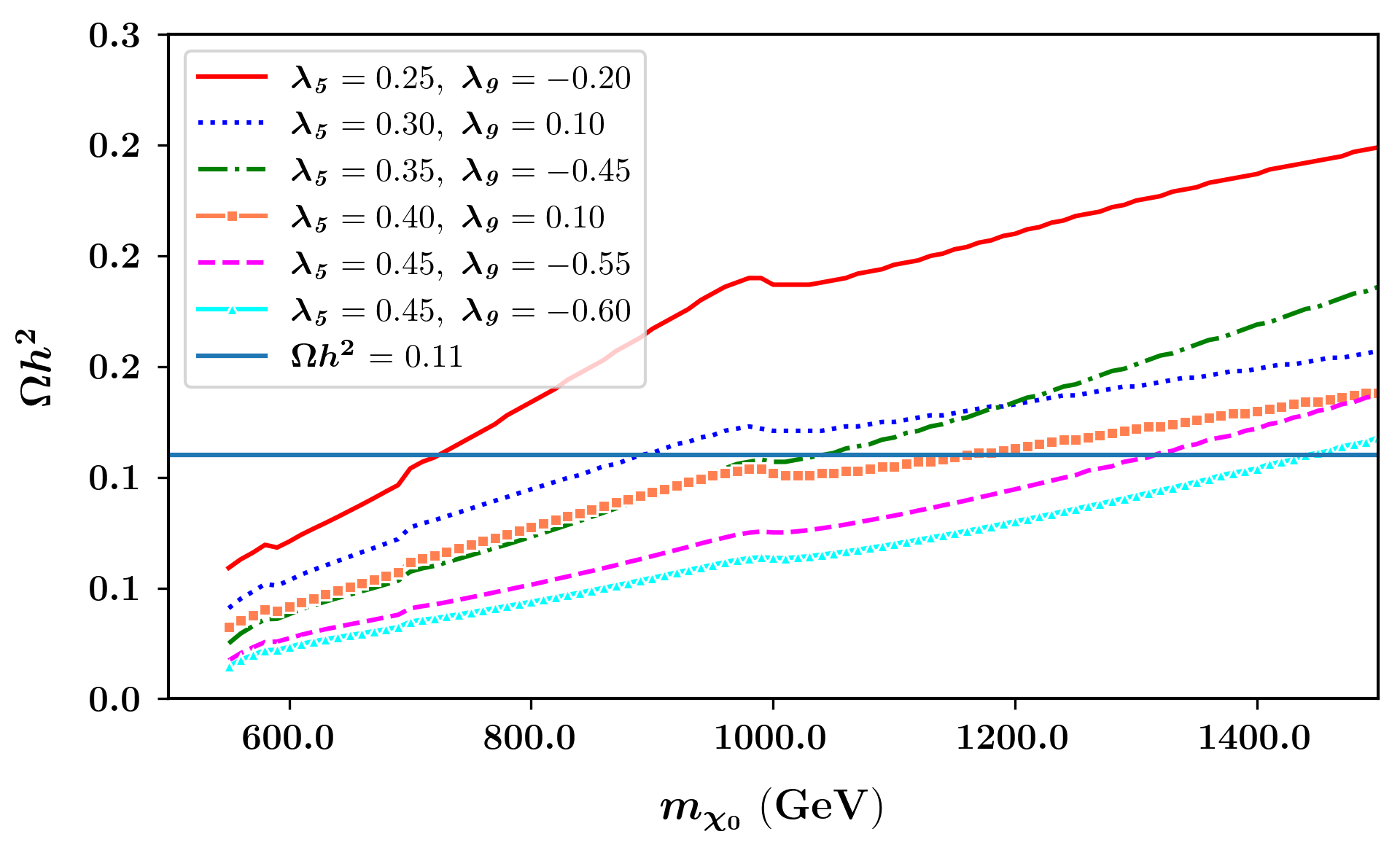}
        \includegraphics[width=0.47\textwidth]{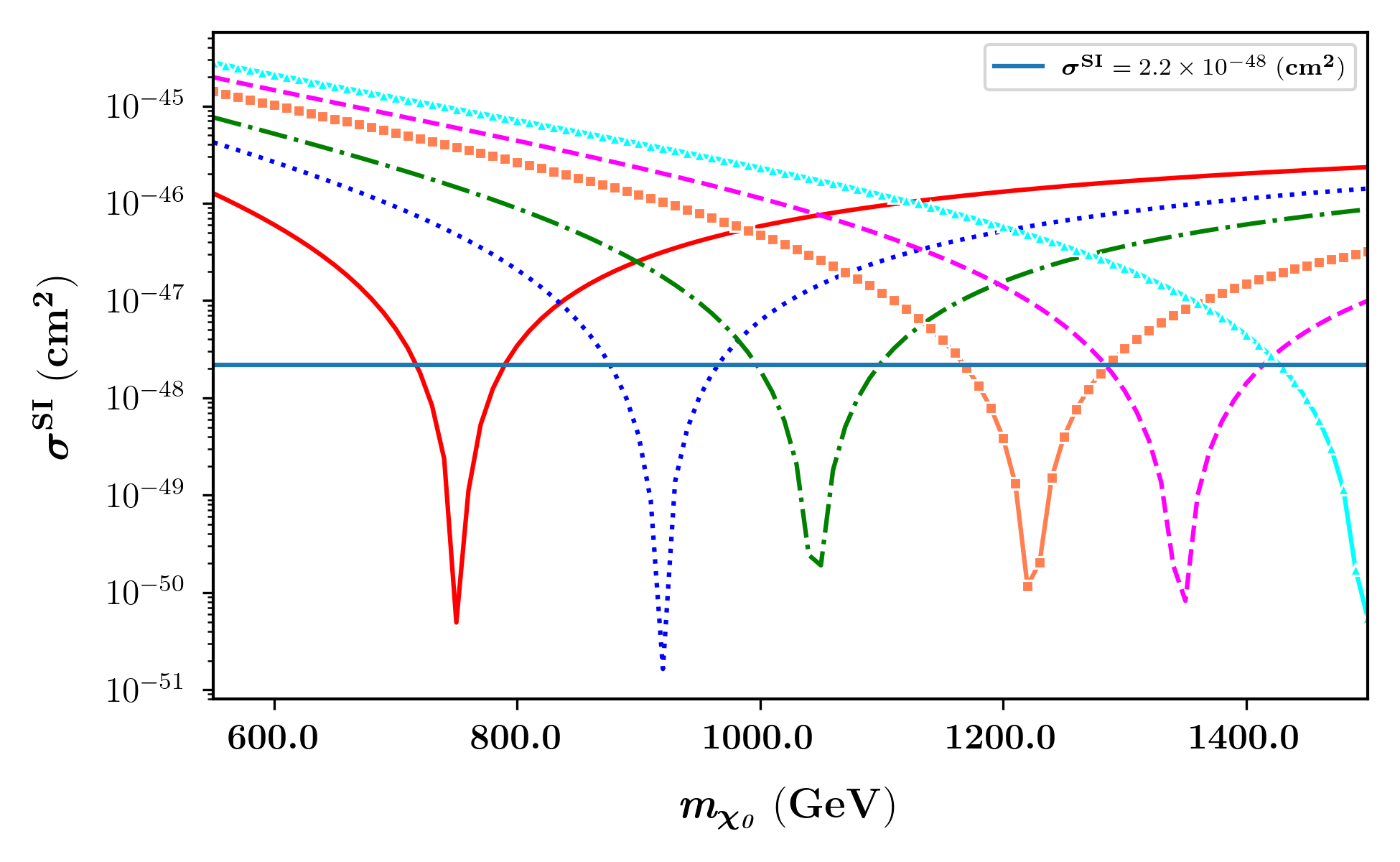}
\caption{Figure presenting the parameter space allowed by relic and blind spot in type-I scenario for $\tan\beta$ =1. The upper left panel shows the allowed parameter space for relic and blind spot in $\lambda_5-\lambda_9$ space, whereas the right upper panel shows the respective allowed parameter space in $\lambda_9-m_{\chi_0}$ plane. The left lower panel shows the variation of relic as function of DM mass (in GeV) for some specific couplings, whereas the lower right panel depicts the direct detection cross-section (in ${\rm cm}^{2}$) as function of DM mass for the same set of points.}
        \label{fig2}
    \end{center}
\end{figure}

We now move on to the discussion on the DM model parameter space consistent with relic abundance and blind spot condition along with the scanned model parameter space obtained from the stability and unitarity limits. From Fig.~\ref{fig1}, we notice that only small mass spitting $\Delta M=m_{H^\pm}-m_H$ (GeV) is allowed with the scanned ranges of $\tan\beta$ values. Therefore, we further restrict ourselves with the parameter space of Eq.~(\ref{param}) to $\Delta M=-1$ GeV and solve for the allowed region that satisfy relic abundance and blind spot for null detection of dark matter for a particular value of $\tan\beta$. In Fig.~\ref{fig2} (in the upper left panel) we present the $\lambda_5-\lambda_9$ parameter space imposing the bounds from DM relic density and the null detection along with the limits obtained from stability and unitarity for $\Delta M=-1$ GeV and $\tan\beta=1$, in type-I framework. We observe a significant range of parameter space allowed by DM abundance is discarded by stability and unitarity limits, the region satisfying the condition for blind spot (Eq.~(\ref{blind})) is completely in agreement with the allowed range of parameters. The same allowed range of parameter is present in $m_{\chi_0}-\lambda_9$ plane as shown in the upper right panel of Fig.~\ref{fig2}. In lower panels of Fig.~\ref{fig2}, we show the variation of DM relic density and direct detection cross-section against dark mass for a few set of parameter points.

\noindent 
It is to be noted that although $\lambda_5<0$ is also allowed by dark matter-Higgs relic abundance criteria, the blind spot condition can be satisfied with positive values of $\lambda_5$ only. The dark matter coupling parameter with Higgs for $\lambda_5=\lambda_6$ is $\lambda_h\propto \left(\frac{\lambda_5}{2}-\frac{m^2_{\chi^{\pm}}-m^2_{\chi_0}}{v^2}\right)$. Therefore, for $\lambda_5<0$, the Higgs-DM coupling $\lambda_h<0$ and as $\lambda_5$ is increased a resonance is reached and $\lambda_h>0$ is achieved. This effect is clearly demonstrated in the lower left panel of Fig.~\ref{fig2}. For increased value of $\lambda_5$, the region of resonance shifts towards the higher values as with the increase of the value of $m_{\chi_0}$ the corresponding difference between the mass squared of the dark particles increases, which is clearly evident from the Eq.~(\ref{blind}).

\begin{figure}[h!tbp]
    \begin{center}
        \includegraphics[width=0.45\textwidth]{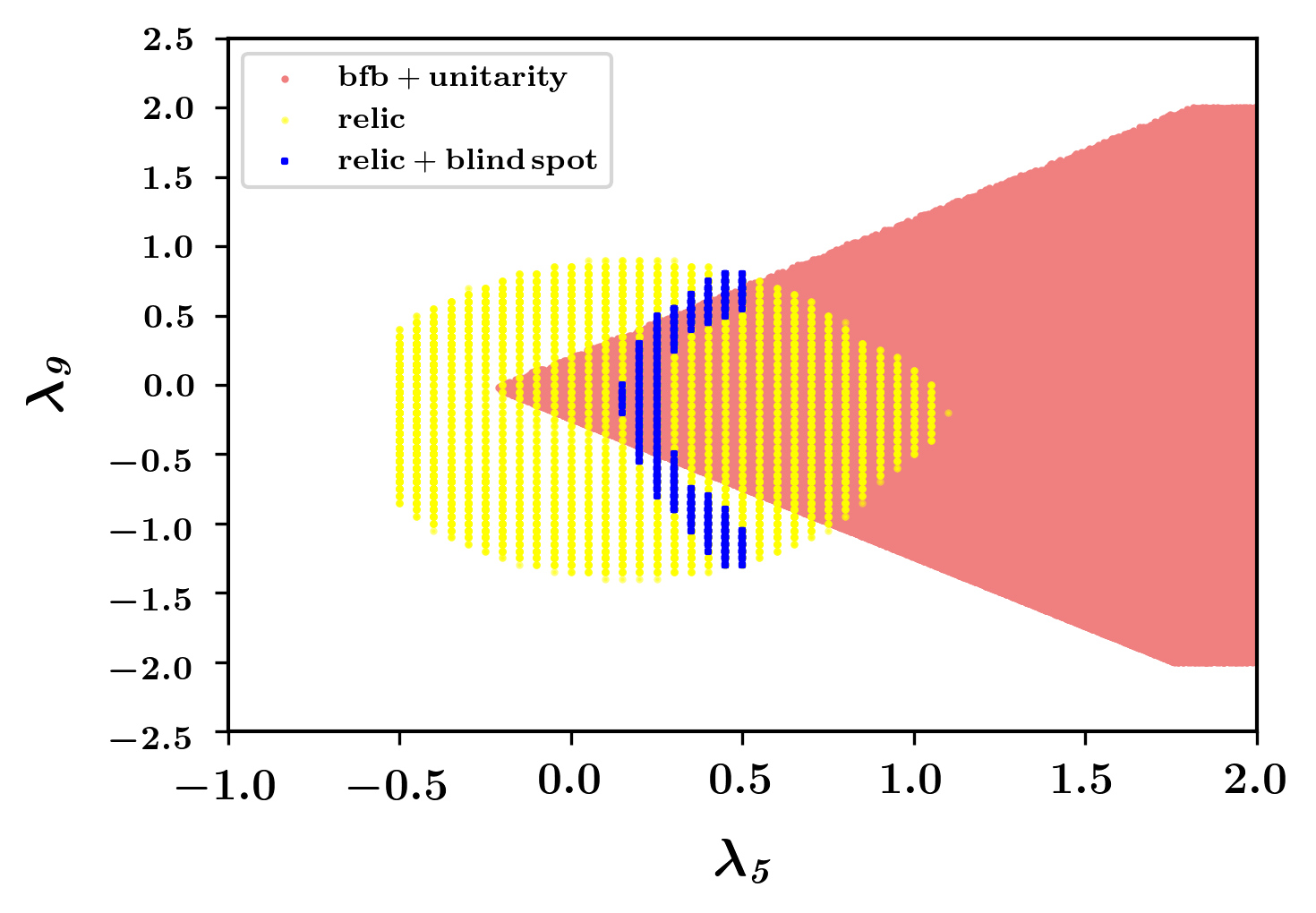}
        \includegraphics[width=0.45\textwidth]{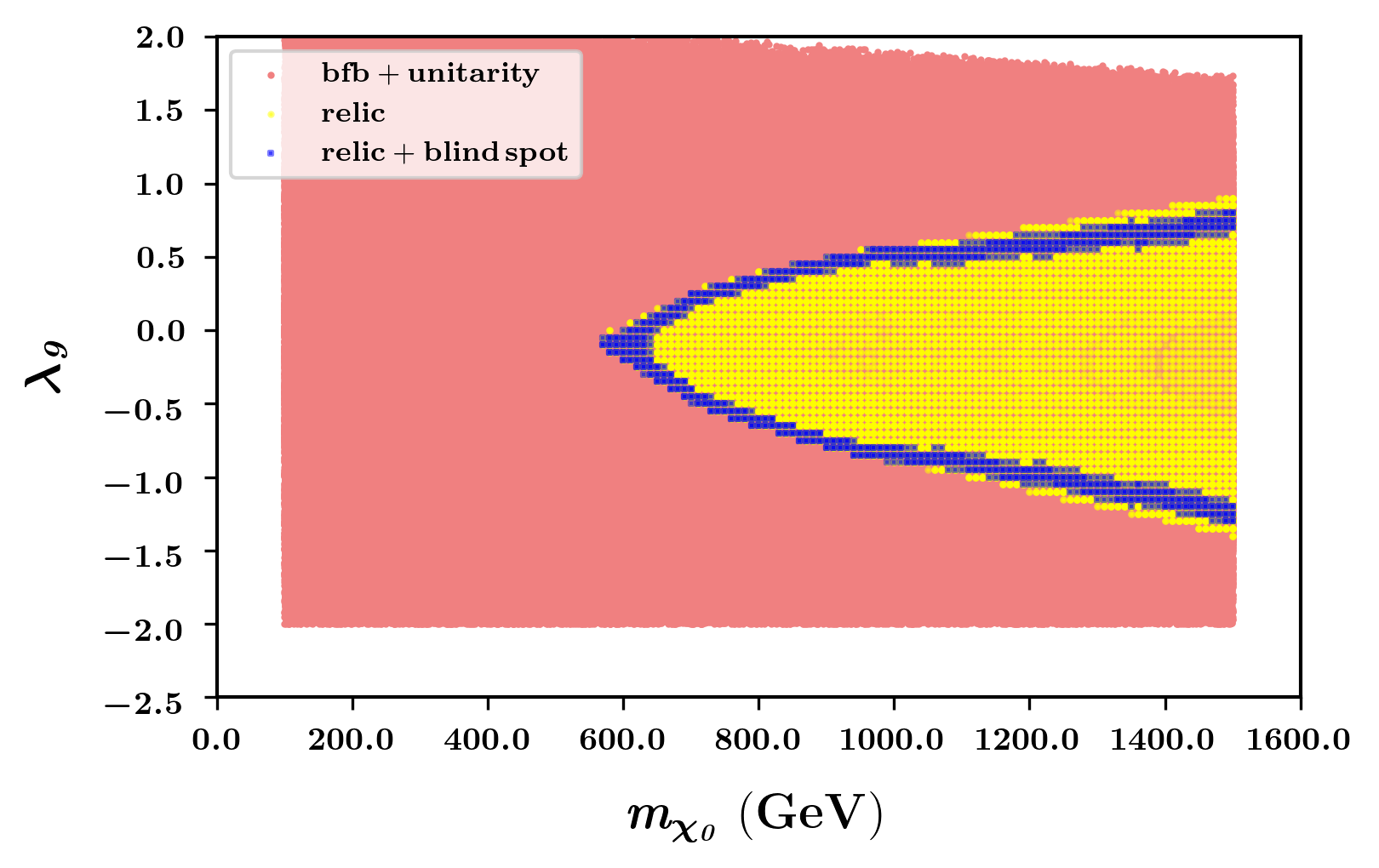}
        \includegraphics[width=0.45\textwidth]{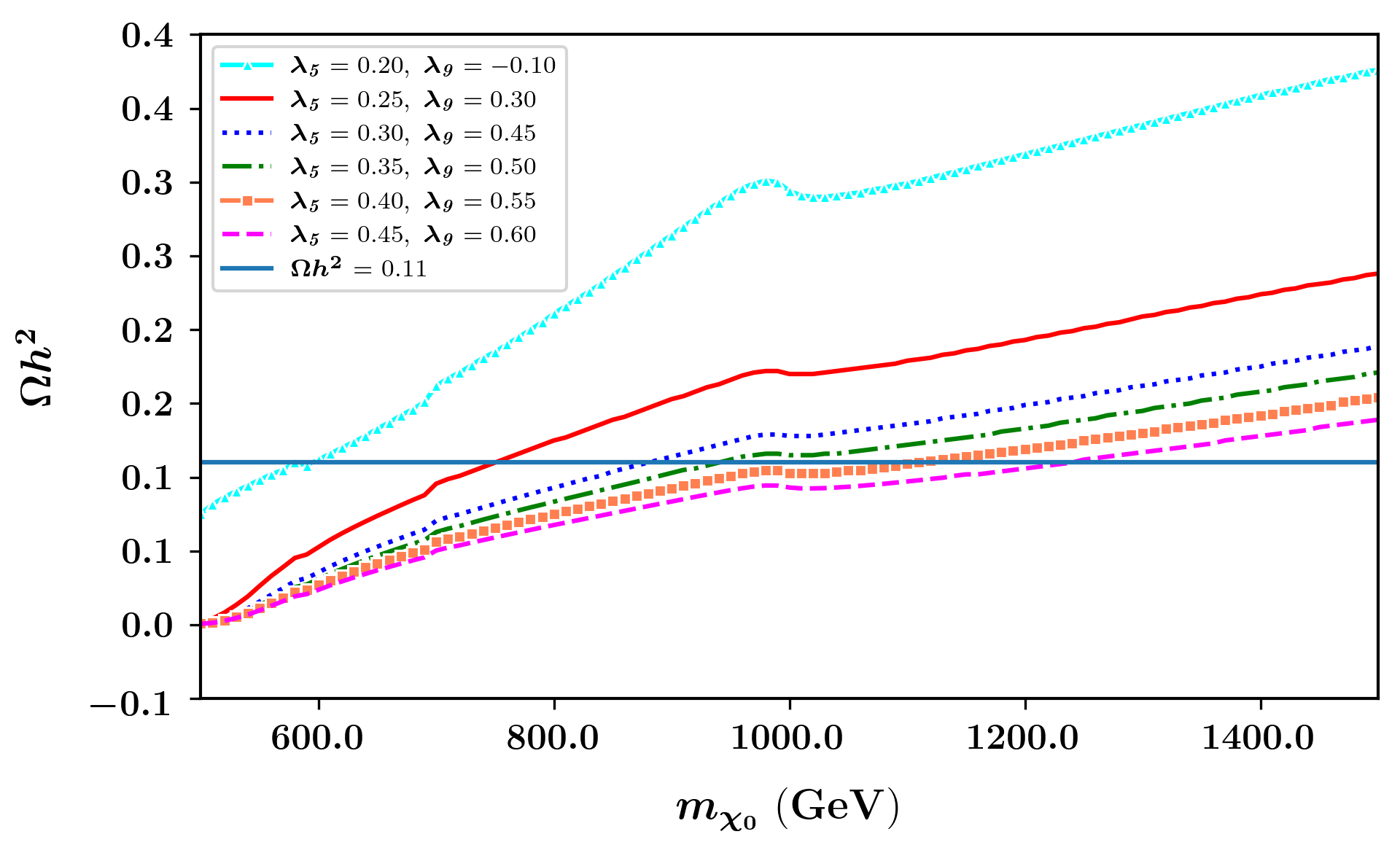}
        \includegraphics[width=0.45\textwidth]{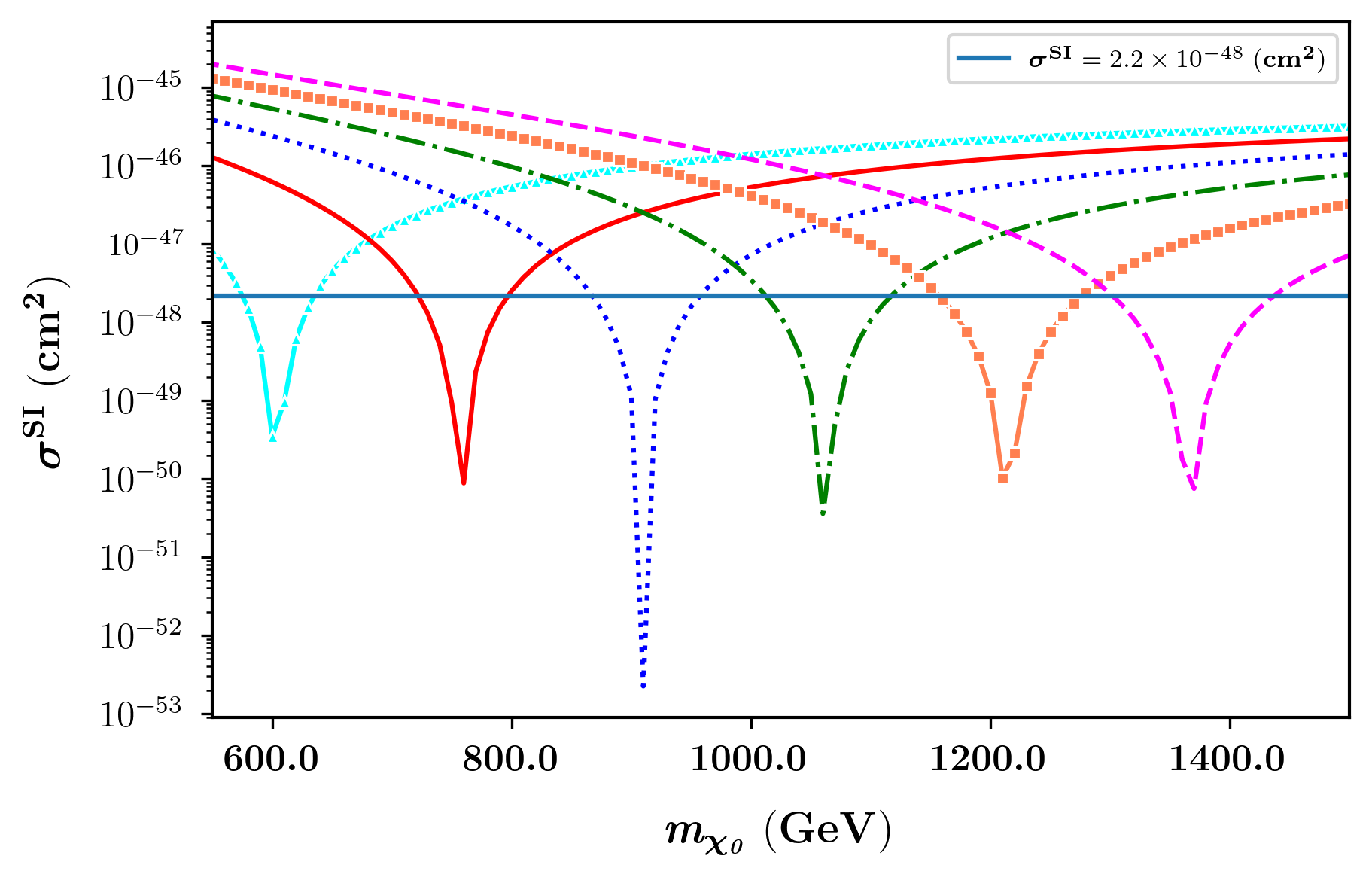}
            \caption{Plots depicting the parameter space allowed by relic and blind spot in type-II scenario for $\tan\beta$ =1 and, $\Delta M=-1$ GeV. The upper panel shows the allowed parameter space for relic and blind spot in $\lambda_5-\lambda_9$ space (left) and in $\lambda_9-m_{\chi_0}$ (GeV) plane (right). The lower panel shows the variation of relic (left) and the direct detection cross-section in ${\rm cm}^{2}$ (right) as function of DM mass (in GeV) for some specific coupling set.}
        \label{fig3}
    \end{center}
\end{figure}

In Fig.~\ref{fig3}, we repeat the same analysis for type-II considering $\Delta M=-1$ GeV and $\tan\beta=1$ for a different set of coupling values. We observe a similar nature of plots in Fig.~\ref{fig3} for type-II framework when compared with Fig.~\ref{fig2}. However, for type-II model, we find the values DM relic density and blind spot condition further restrict the allowed parameter space within the range $|\lambda_9|\leq 0.5$. Therefore,  for $\tan\beta=1$, a significant range of model parameter space with 
with $m_{\chi_0}\geq700$ GeV satisfying $\lambda_9\leq - 0.5$ (lower segment of $m_{\chi_0}-\lambda_9$ region in the plot) 
is discarded. Similarly a small range of parametere space with $m_{\chi_0}\geq 1400$  GeV  gets exclueded for $\lambda_9\geq 0.5$. 
The variation of relic density and direct detection cross-section plotted against dark matter mass is depicted in the lower panel of Fig.~\ref{fig3} for a few set of points. For both type-I and type-II two Higgs doublet scenario considered, we observe significant decrease in the available model parameter space when relic density and direct detection (blind spot criteria) is imposed along with the stability and unitarity limits. For larger $\tan\beta$ values, the DM relic density gets suppressed significantly for dark matter mass range considered $500~{\rm GeV}\leq m_{\chi_0}\leq 1500$ GeV. 
The inert dark matter annihilates mostly into gauge bosons for $\tan\beta=1$. With increase in $\tan\beta$, DM annihilation 
$\chi_0\chi_0\rightarrow hH$ and co-annihilation $\chi^+\chi_0\rightarrow hH^+$ dominate and reduction in the value of relic abundance is observed for both type-I and type-II THDM framework for same value of coupling parameters, thus pushing the relic satisfied region shifting towards higher DM mass. Additionally, for type-I model DM annihilations $\chi_0\chi^+\rightarrow W^+ H$ and $\chi_A\chi^+\rightarrow Z H^+$ channels also contribute significantly resulting further reduction of relic density. For heavier DM mass above 1 TeV, contributions from new annihilation channels $\chi_0\chi_0\rightarrow HH,AA$, and $\chi_0\chi^+\rightarrow HH^+$ further reduce the relic abundance significantly. 
With careful scrutiny, we put conservative bounds ruling out $\tan\beta\geq 5$ for type-I THDM framework and $\tan\beta\geq 10$ for type-II THDM model. The limit on $\tan\beta$ can be relaxed by considering a higher dark matter mass range above 1.5 TeV.

\begin{figure}[h!tbp]
    \begin{center}
  \hspace*{-2cm}      \includegraphics[width=0.5\textwidth]{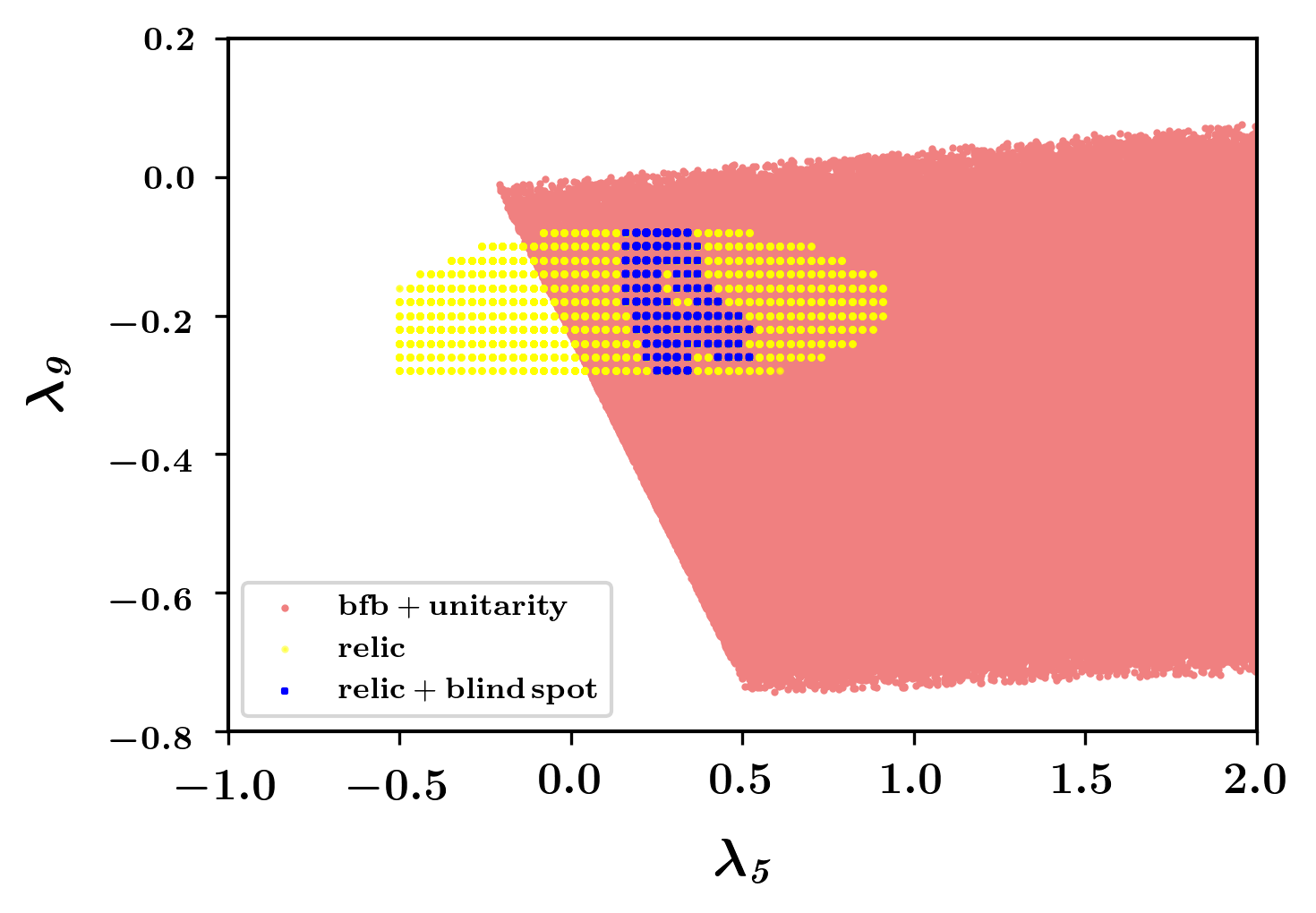}
        \includegraphics[width=0.5\textwidth]{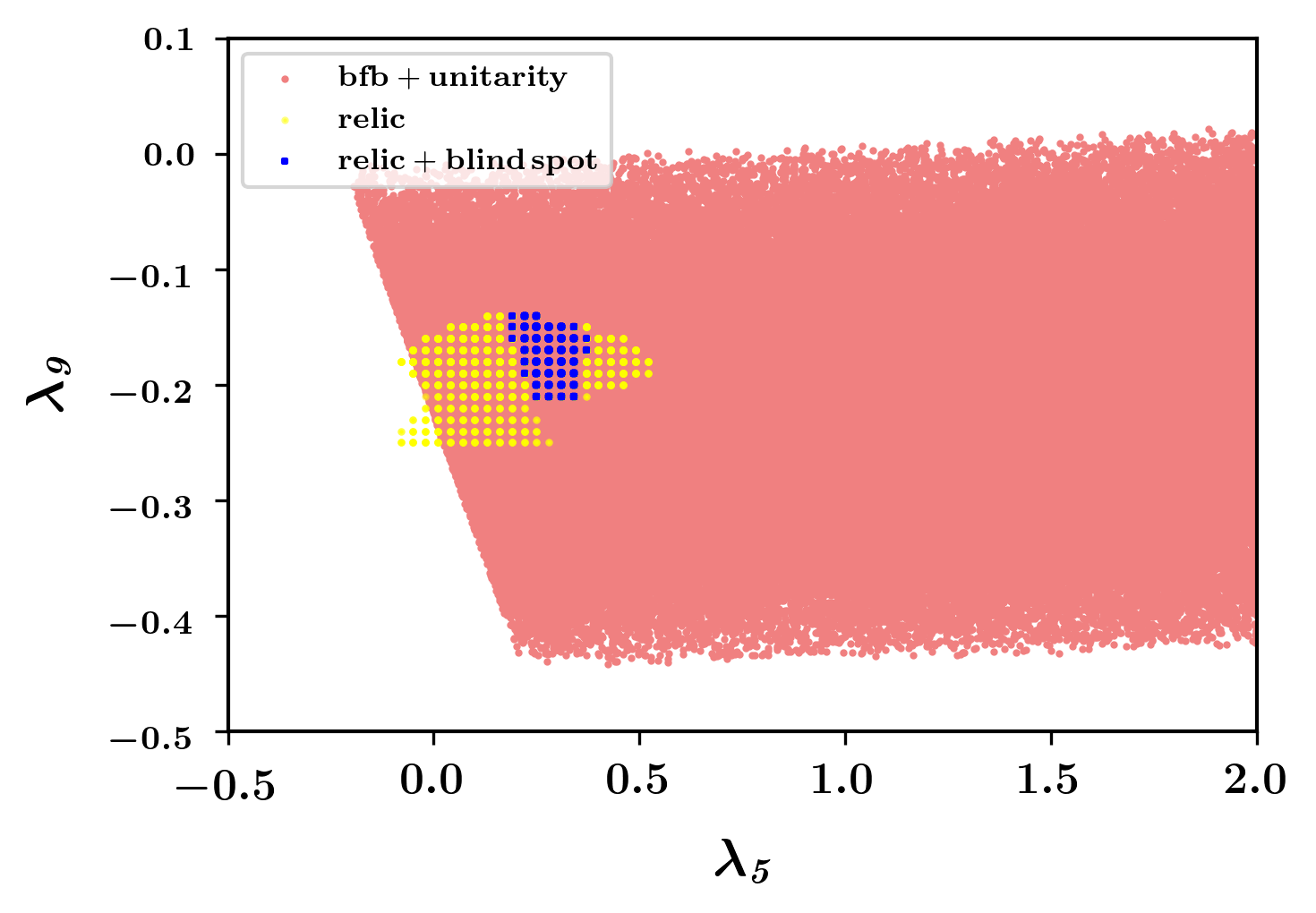}
  \hspace*{-2cm}      \includegraphics[width=0.5\textwidth]{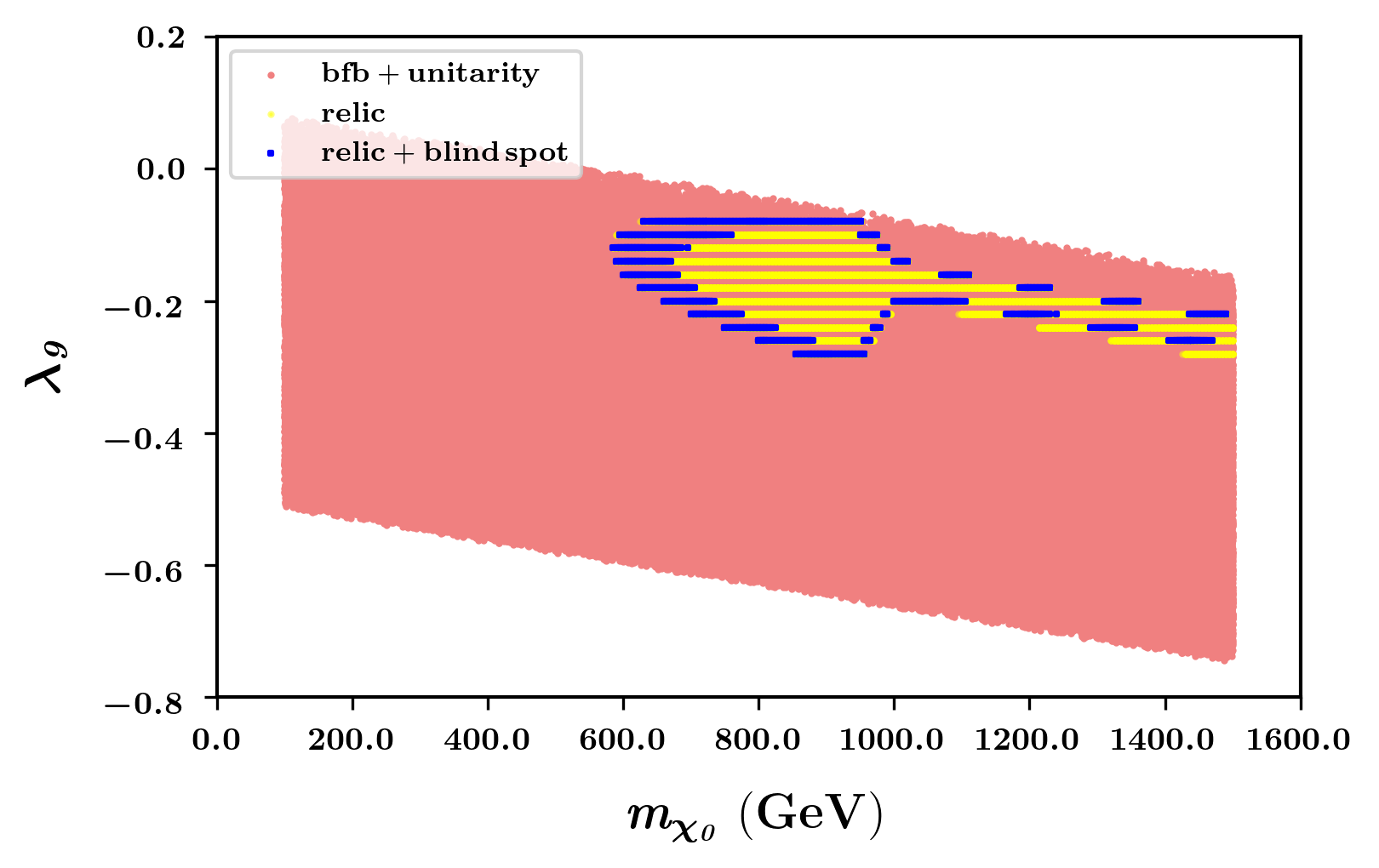}
        \includegraphics[width=0.5\textwidth]{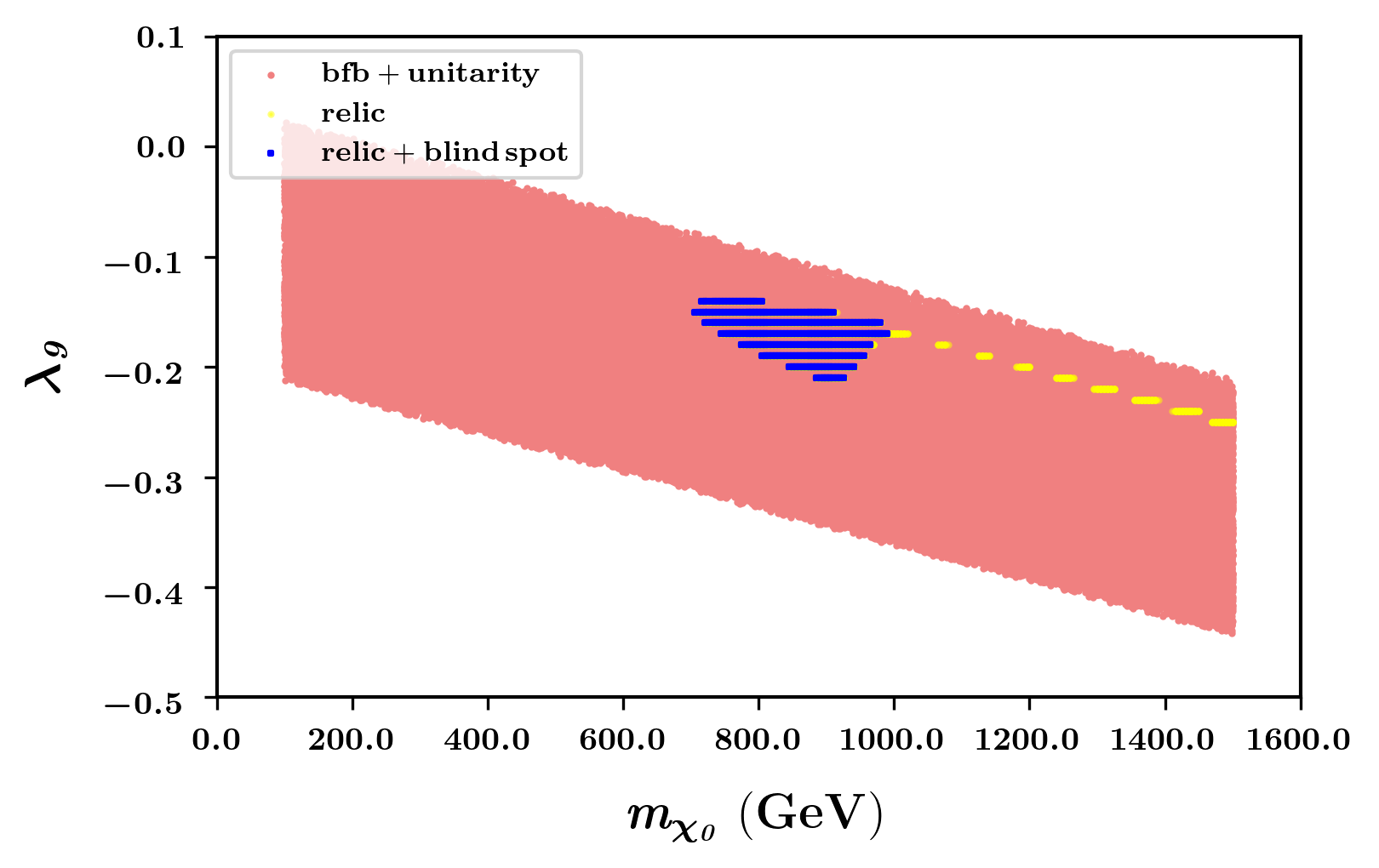}
  \hspace*{-2cm}      \includegraphics[width=0.5\textwidth]{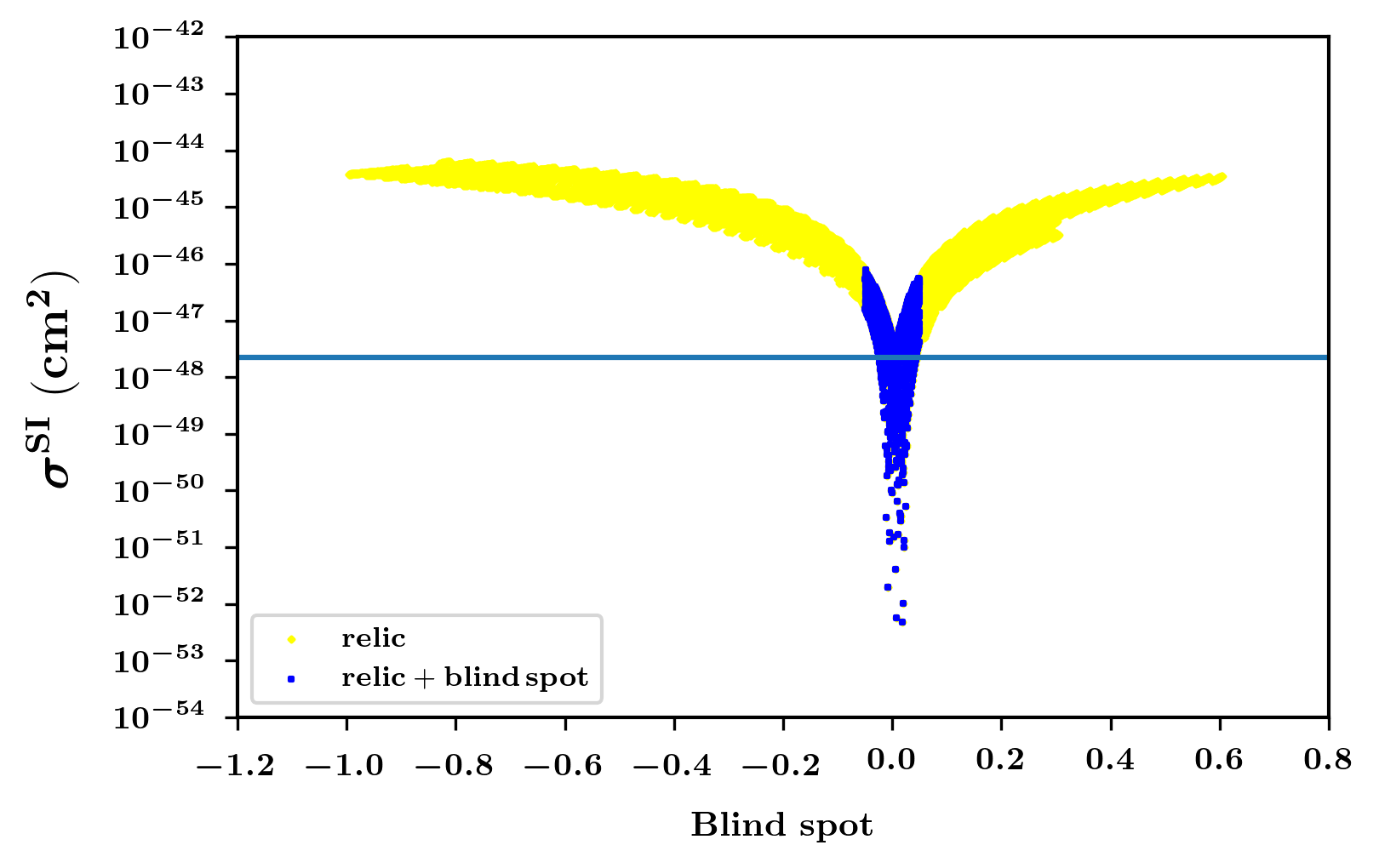}
        \includegraphics[width=0.5\textwidth]{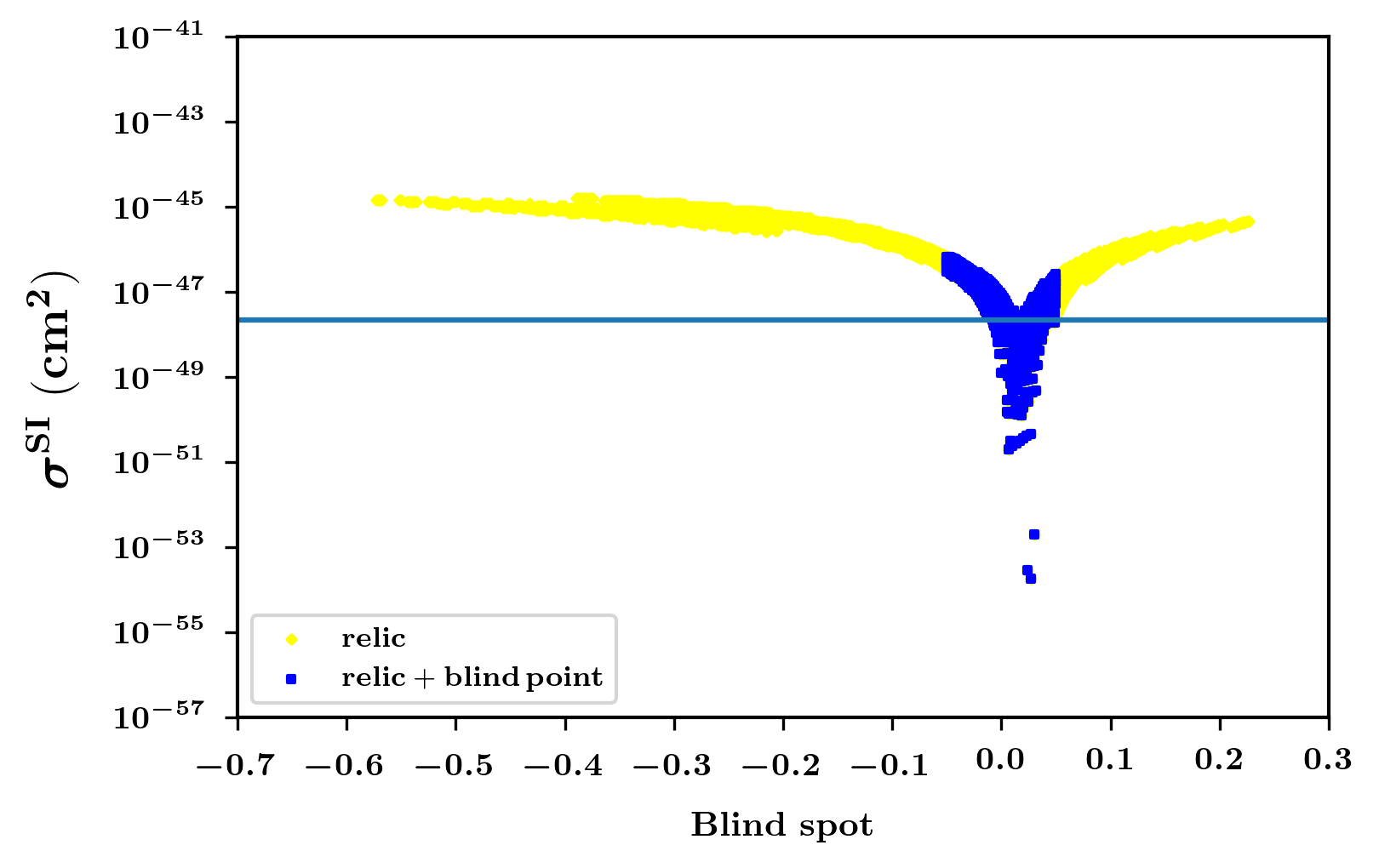}
            \caption{Plots showing the parameter space allowed by relic and blind spot in type-II scenario for $\tan\beta$ =5 (left panel) and $\tan\beta$ =8 (right panel) for $\Delta M=-1$ GeV. The upper most panel and the middle one present the allowed parameter space in $\lambda_9 - \lambda_5$ space and in $\lambda_9-m_{\chi_0}$ (GeV) plane. The last one shows the variation of SI cross-section (in ${\rm cm}^{2}$) as function of blind spot.}
        \label{fig4}
    \end{center}
\end{figure}

In Fig.~\ref{fig4}, we demonstrate the effect of increasing $\tan\beta$ on the allowed dark matter model parameter space for type-II THDM framework and compare the results for $\tan\beta=5$ and $\tan\beta=8$. It is easily observed from the upper panel of Fig.~\ref{fig4}, that the allowed $\lambda_5-\lambda_9$ plane shrinks with large $\tan\beta$. The same conclusion can be drawn from the $\lambda_9-m_{\chi_0}$ parameter space shown in the middle panel of Fig.~\ref{fig4}. Interestingly, with increased $\tan\beta$, the region of allowed dark matter parameter space that satisfies blind spot condition narrows down towards low mass dark matter regime. A similar effect is also observed for the model parameter space with type-I.  Finally, in the lower panel, we plot the dark matter scattering cross-section with the condition for blind spot (Eq.~(\ref{blind})) with $\tan\beta=5$ and $\tan\beta=8$. This confirms that $\lambda_h\rightarrow 0$ as we approach the condition to achieve blind spot resulting $\sigma_{\rm SI}^h \rightarrow 0$.

\subsection{Statistical analysis of model parameters}
\label{subsec:PLR}

We now investigate the statistical weight of the viable model parameter space. We consider a frequentist method \cite{Storn:1997uea,AbdusSalam:2020rdj} with three model parameters $\lambda_5,~\lambda_9$, and $m_{\chi_0}$ and calculate the likelihood function for the model
using constraints from the relic density and the direct detection experiment. The likelihood function in the present formalism is given as \cite{Ellis:2017ndg,Yaguna:2024jor}

\begin{align}
    \mathcal{L}(\vec{x})=\mathcal{L}_{\Omega}(\vec{x})\times\mathcal{L}_\text{DD}(\vec{x})\, .
\end{align}
The likelihood function is maximized using the Wilk's theorem \cite{Wilks:1938dza} to obtain the 1$\sigma$ and 2$\sigma$ confidence level (C.L.) regions of the model parameters. We scrutinize the model parameters by the profile likelihood ratio (PLR)
\begin{align}
   {{\rm PLR}({\mathcal L}(x_k))}=\frac{\mathcal{L}(x_k)}{\mathcal{L}_{\rm max}},
\end{align}
where 
\begin{align}
    &\mathcal{L}(x_k)=\underset{x_k\notin \vec{x}}{{\rm  max}}\,[\mathcal{L}(\vec{x})],\hspace{1cm}\mathcal{L}_{\rm max}={\rm max}\,[\mathcal{L}(\vec{x})]\, .
\end{align}
The likelihood functions for any observable are Gaussian and can be expressed as
\begin{align}
    \chi_\mathcal{Q}^2(\vec{x})=-2\ln \mathcal{L}_\mathcal{Q}(\vec{x})=\left(\frac{\mathcal{Q}^{\text{th}}-\mathcal{Q}^{\text{obs}}}{\Sigma_\mathcal{Q}}\right)^2, 
\end{align}
where  $\mathcal{Q}^{\text{th}}$ and $\mathcal{Q}^{\text{obs}}$ specify the theoretical and experimental values of observable quantity $\mathcal{Q}=\Omega,{\rm DD}$, and $\Sigma_\mathcal{Q}$ is the standard deviation. We adopt the formalism of ref.~\cite{Yaguna:2024jor} and consider $\Omega^{\text{th}}$ having values between 0.105 to 0.125 
and $\Sigma^2_{\sigma_\text{SI}}=[0.2\sigma_\text{SI}^\text{h}]^2 + [\Sigma_{\sigma_\text{SI}}^\text{LZ}/1.64]^2$, assuming a 20\% uncertainty in the theoretical values obtained and 90\% C. L. upper limit on $\sigma_{\rm SI}$ from LZ experiment \cite{LZ:2022lsv}.

\begin{figure}[h!tbp]
    \begin{center}
        \includegraphics[width=0.45\textwidth]{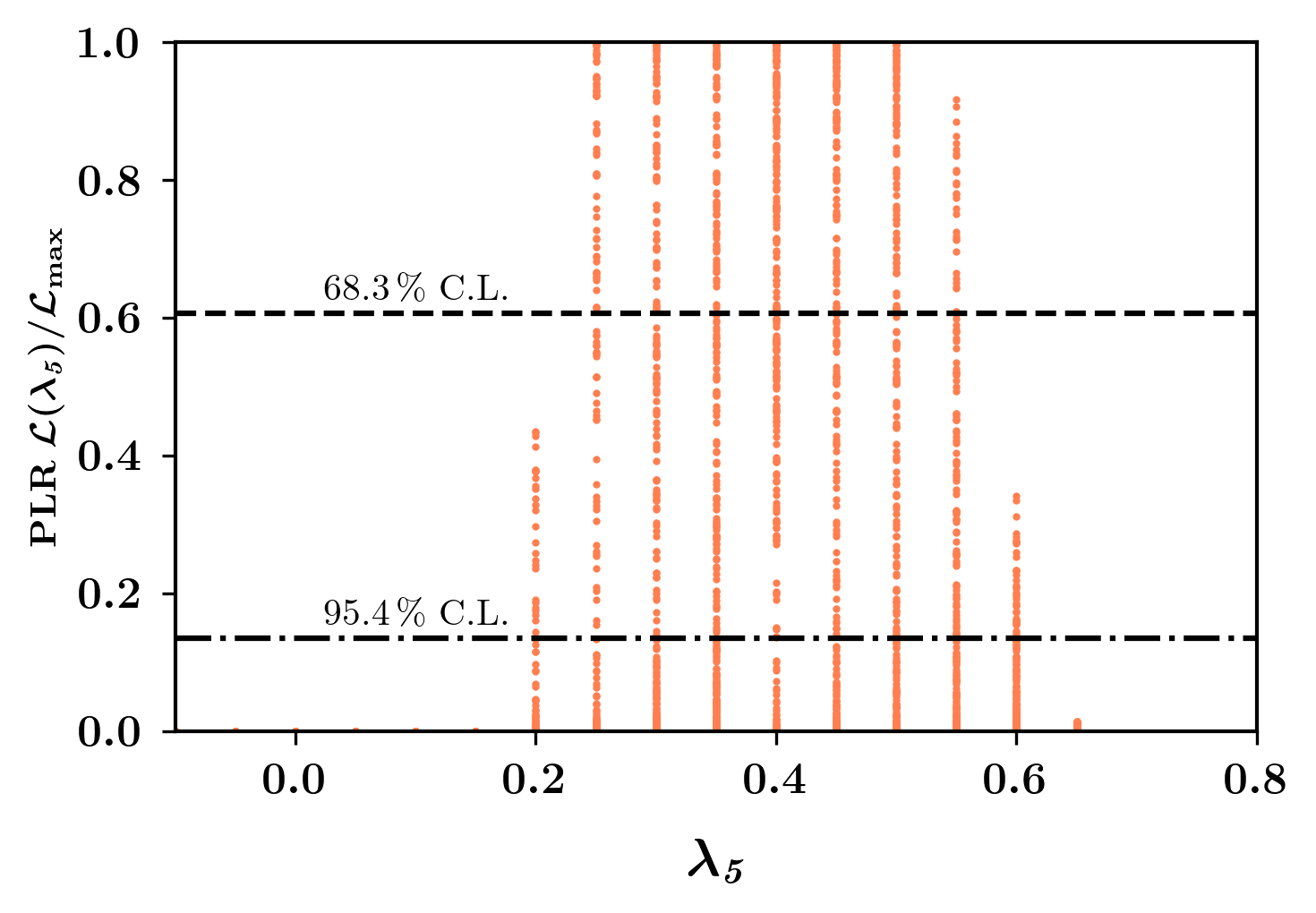}
        \includegraphics[width=0.45\textwidth]{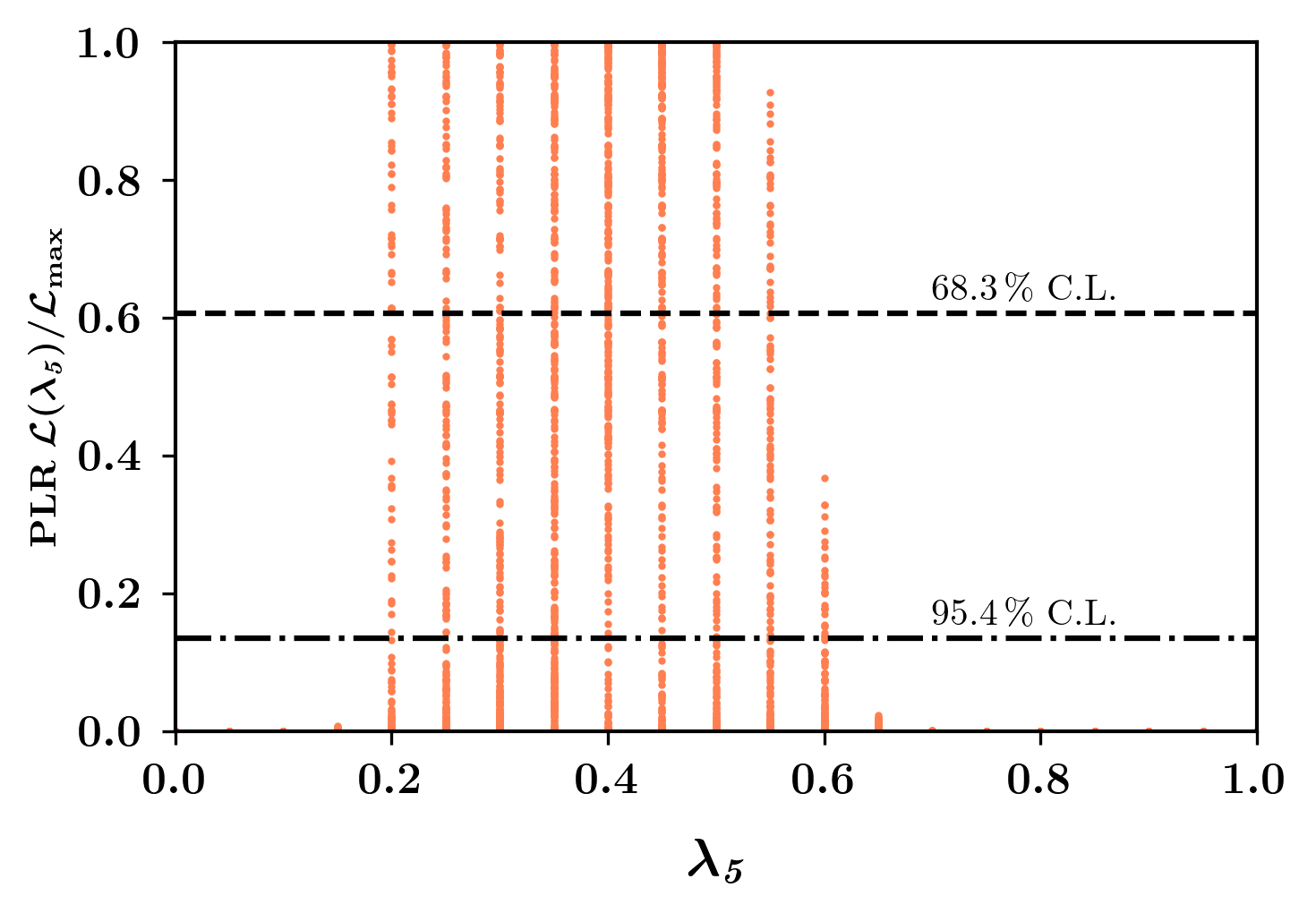}\\
        \includegraphics[width=0.45\textwidth]{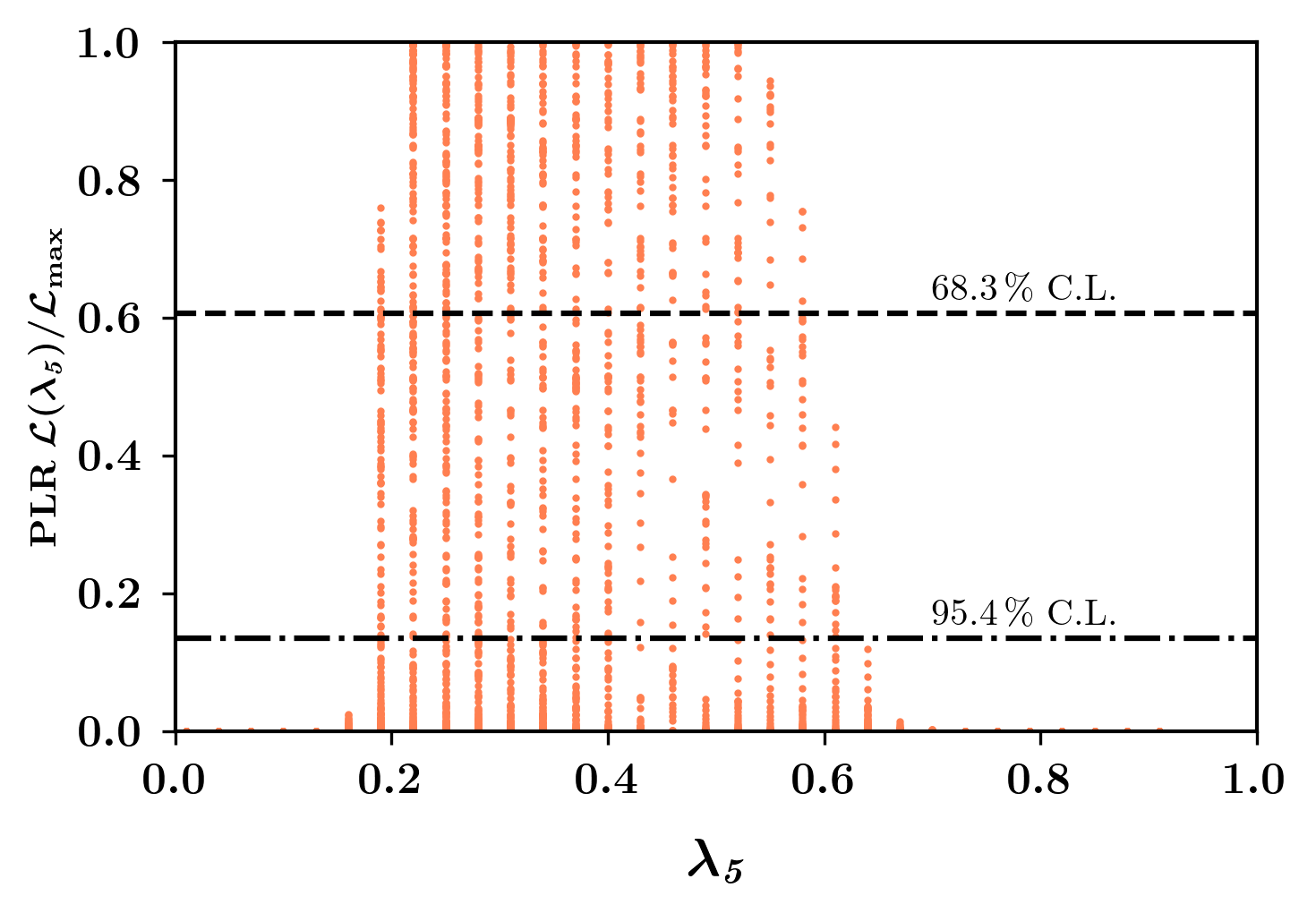}
         \includegraphics[width=0.45\textwidth]{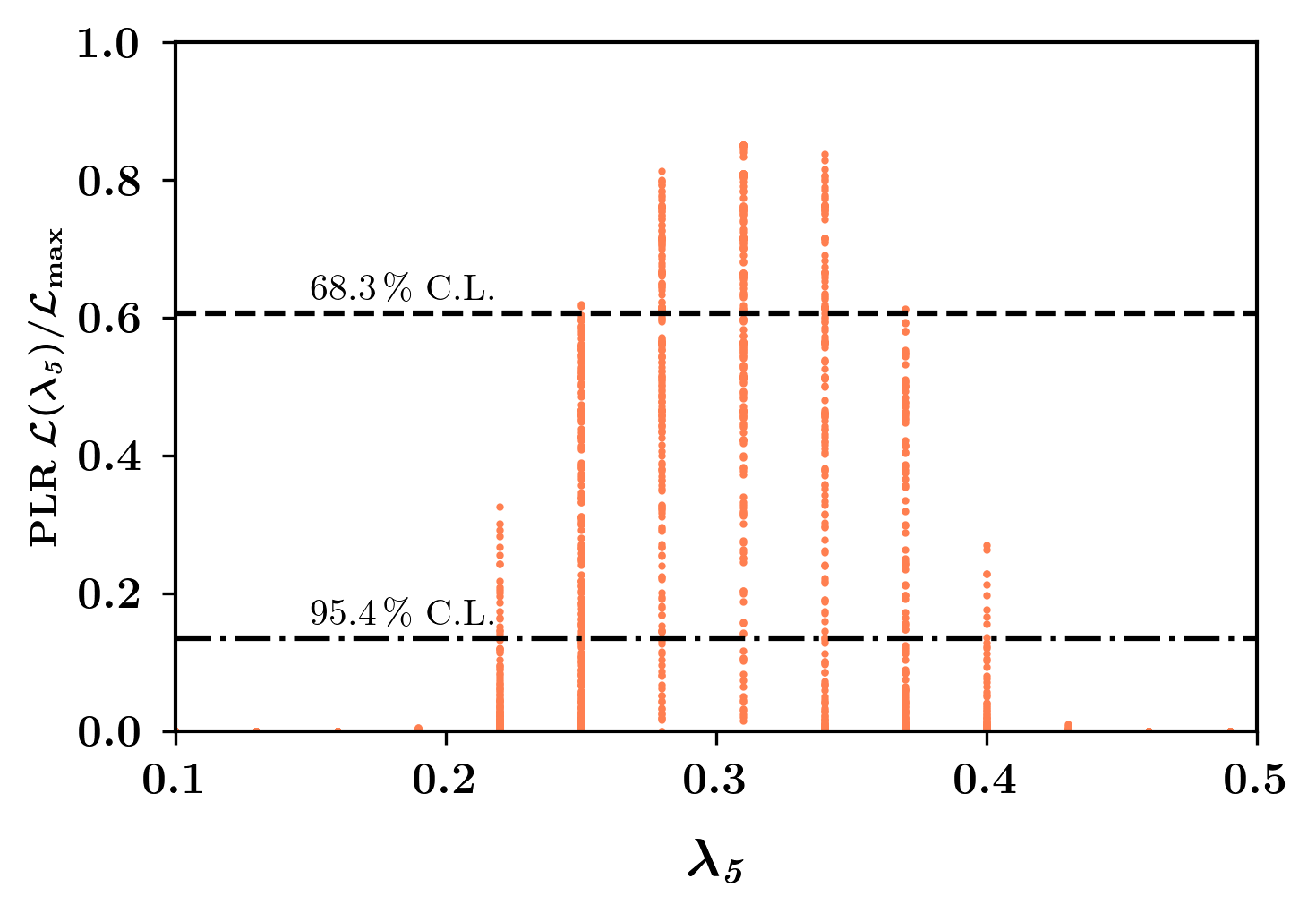}
            \caption{Upper left (right) panel shows 1 dimensional PLR of parameter $\lambda_5$ plotted with  $\tan\beta=1$ for inert DM in type-I (type-II) model. Lower left (right) panel shows the same for inert DM in type-II model using $\tan\beta=5$ ($\tan\beta=8$).}
        \label{fig5}
    \end{center}
\end{figure}

In Fig.~\ref{fig5}, we demonstrate the PLR for the coupling parameter $\lambda_5$ with the available model parameter space  for both type-I and type-II frameworks obtained from Fig.~\ref{fig2} and Figs.~\ref{fig3}-\ref{fig4}. From the 1 dimensional PLR plot for $\lambda_5$, the $1\sigma$ ($2\sigma$) C.L. region can be estimated by the relation $-2\ln\left(\mathcal{L}(\lambda_5)/\mathcal{L}_{{\rm max}}\right)\leq 1\,(4)$. 
From Fig.~\ref{fig5}, it can be clearly noticed that from global fitting
for type-I scenario $0.25\leq \lambda_5\leq 0.55$ is favoured for $\tan\beta=1$ (upper left panel); whereas for type-II, for same $\tan\beta$ the range $0.2\leq \lambda_5\leq 0.55$ is favoured.
The same can be concluded for $\tan\beta=5$ for type-II THDM. However the allowed region shrinks further between $0.25\leq \lambda_5\leq 0.35$ for $\tan\beta=8$.
The result is also in agreement with the earlier findings leading to
reduction in available model parameter space for an increase in $\tan\beta$ value.

\section{Conclusions}
\label{sec:conclusion}

In this work, we address the plausible scenario of inert doublet dark matter embedded with two Higgs doublet model resulting a three Higgs doublet model with two active and one inert doublet. The inert dark matter in the three Higgs doublet featured with type-I and type-II two Higgs doublet framework, has an unique characteristics of null detection resulting a blind spot for dark matter detection. Applying the stability and unitarity limits on both the models 
impose tight limit on the mass splitting of heavy scalar of two Higgs doublets, resulting a narrow range of model parameters allowed for large range of $\tan\beta$ values within the alignment limit of two Higgs doublet. Neglecting the contribution of heavy scalar to the scattering amplitude for dark matter direct detection leads to a simple condition of $\lambda_5\frac{v^2}{2}=(m^2_{\chi^{\pm}}-m^2_{\chi_0})$ for blind spot of direct detection. Further constraints imposed on model parameter space consistent with relic abundance of DM, direct detection bounds reveals that the type-I framework is less favoured with the increased $\tan\beta$ due to increased annihilation cross-section. A similar result is obtained for the inert doublet dark matter with type-II two Higgs doublet with a relaxed parameter space allowing values of $\tan\beta\leq 10$ for the dark matter mass ranging below 1.5 TeV. 
We further investigate the statistical significance of the model parameter. 
The profile likelihood study of the model parameter also confirms the less availability of inert dark matter
model parameters with enhancement in $\tan\beta$ values for both types of THDM considered. With the increased sensitivity of direct detection experiments, the blind spot narrative of dark matter provides a feasible solution towards no observation of dark matter signal. This leads to complementary probes of the model through collider physics or indirect searches, can be pursued in future works.

\section*{Acknowledgment} The authors acknowledge Ipsita Saha, \`{O}scar Zapata, Carlos E. Yaguna and Debarun Paul for useful discussions. ADB acknowledges financial support from DST, India, under grant number IFA20-PH250 (INSPIRE Faculty Award). TJ acknowledges the projects Geracles-hanke, C-SpaRC-hanke for financial support.

\bibliographystyle{JHEP}
\bibliography{reference}

\providecommand{\href}[2]{#2}\begingroup\raggedright\begin{thebibliography}{10}

\bibitem{Planck:2018vyg}
{\scshape Planck} collaboration, \emph{{Planck 2018 results. VI. Cosmological
  parameters}},
  \href{https://doi.org/10.1051/0004-6361/201833910}{\emph{Astron. Astrophys.}
  {\bfseries 641} (2020) A6}
  [\href{https://arxiv.org/abs/1807.06209}{{\ttfamily 1807.06209}}].

\bibitem{Jungman:1995df}
G.~Jungman, M.~Kamionkowski and K.~Griest, \emph{{Supersymmetric dark matter}},
  \href{https://doi.org/10.1016/0370-1573(95)00058-5}{\emph{Phys. Rept.}
  {\bfseries 267} (1996) 195}
  [\href{https://arxiv.org/abs/hep-ph/9506380}{{\ttfamily hep-ph/9506380}}].

\bibitem{Bertone:2004pz}
G.~Bertone, D.~Hooper and J.~Silk, \emph{{Particle dark matter: Evidence,
  candidates and constraints}},
  \href{https://doi.org/10.1016/j.physrep.2004.08.031}{\emph{Phys. Rept.}
  {\bfseries 405} (2005) 279}
  [\href{https://arxiv.org/abs/hep-ph/0404175}{{\ttfamily hep-ph/0404175}}].

\bibitem{CMS:2012qbp}
{\scshape CMS} collaboration, \emph{{Observation of a New Boson at a Mass of
  125 GeV with the CMS Experiment at the LHC}},
  \href{https://doi.org/10.1016/j.physletb.2012.08.021}{\emph{Phys. Lett. B}
  {\bfseries 716} (2012) 30} [\href{https://arxiv.org/abs/1207.7235}{{\ttfamily
  1207.7235}}].

\bibitem{ATLAS:2012yve}
{\scshape ATLAS} collaboration, \emph{{Observation of a new particle in the
  search for the Standard Model Higgs boson with the ATLAS detector at the
  LHC}}, \href{https://doi.org/10.1016/j.physletb.2012.08.020}{\emph{Phys.
  Lett. B} {\bfseries 716} (2012) 1}
  [\href{https://arxiv.org/abs/1207.7214}{{\ttfamily 1207.7214}}].

\bibitem{Huang:2014xua}
P.~Huang and C.E.M.~Wagner, \emph{{Blind Spots for neutralino Dark Matter in
  the MSSM with an intermediate $m_A$}},
  \href{https://doi.org/10.1103/PhysRevD.90.015018}{\emph{Phys. Rev. D}
  {\bfseries 90} (2014) 015018}
  [\href{https://arxiv.org/abs/1404.0392}{{\ttfamily 1404.0392}}].

\bibitem{OHare:2015utx}
C.A.J.~O'Hare, A.M.~Green, J.~Billard, E.~Figueroa-Feliciano and L.E.~Strigari,
  \emph{{Readout strategies for directional dark matter detection beyond the
  neutrino background}},
  \href{https://doi.org/10.1103/PhysRevD.92.063518}{\emph{Phys. Rev. D}
  {\bfseries 92} (2015) 063518}
  [\href{https://arxiv.org/abs/1505.08061}{{\ttfamily 1505.08061}}].

\bibitem{Crivellin:2015bva}
A.~Crivellin, M.~Hoferichter, M.~Procura and L.C.~Tunstall, \emph{{Light stops,
  blind spots, and isospin violation in the MSSM}},
  \href{https://doi.org/10.1007/JHEP07(2015)129}{\emph{JHEP} {\bfseries 07}
  (2015) 129} [\href{https://arxiv.org/abs/1503.03478}{{\ttfamily
  1503.03478}}].

\bibitem{Badziak:2015exr}
M.~Badziak, M.~Olechowski and P.~Szczerbiak, \emph{{Blind spots for neutralino
  dark matter in the NMSSM}},
  \href{https://doi.org/10.1007/JHEP03(2016)179}{\emph{JHEP} {\bfseries 03}
  (2016) 179} [\href{https://arxiv.org/abs/1512.02472}{{\ttfamily
  1512.02472}}].

\bibitem{Han:2016qtc}
T.~Han, F.~Kling, S.~Su and Y.~Wu, \emph{{Unblinding the dark matter blind
  spots}}, \href{https://doi.org/10.1007/JHEP02(2017)057}{\emph{JHEP}
  {\bfseries 02} (2017) 057}
  [\href{https://arxiv.org/abs/1612.02387}{{\ttfamily 1612.02387}}].

\bibitem{Cheung:2013dua}
C.~Cheung and D.~Sanford, \emph{{Simplified Models of Mixed Dark Matter}},
  \href{https://doi.org/10.1088/1475-7516/2014/02/011}{\emph{JCAP} {\bfseries
  02} (2014) 011} [\href{https://arxiv.org/abs/1311.5896}{{\ttfamily
  1311.5896}}].

\bibitem{Baum:2017enm}
S.~Baum, M.~Carena, N.R.~Shah and C.E.M.~Wagner, \emph{{Higgs portals for
  thermal Dark Matter. EFT perspectives and the NMSSM}},
  \href{https://doi.org/10.1007/JHEP04(2018)069}{\emph{JHEP} {\bfseries 04}
  (2018) 069} [\href{https://arxiv.org/abs/1712.09873}{{\ttfamily
  1712.09873}}].

\bibitem{Dedes:2014hga}
A.~Dedes and D.~Karamitros, \emph{{Doublet-Triplet Fermionic Dark Matter}},
  \href{https://doi.org/10.1103/PhysRevD.89.115002}{\emph{Phys. Rev. D}
  {\bfseries 89} (2014) 115002}
  [\href{https://arxiv.org/abs/1403.7744}{{\ttfamily 1403.7744}}].

\bibitem{Branco:2011iw}
G.C.~Branco, P.M.~Ferreira, L.~Lavoura, M.N.~Rebelo, M.~Sher and J.P.~Silva,
  \emph{{Theory and phenomenology of two-Higgs-doublet models}},
  \href{https://doi.org/10.1016/j.physrep.2012.02.002}{\emph{Phys. Rept.}
  {\bfseries 516} (2012) 1} [\href{https://arxiv.org/abs/1106.0034}{{\ttfamily
  1106.0034}}].

\bibitem{Gunion:1989we}
J.F.~Gunion, H.E.~Haber, G.L.~Kane and S.~Dawson, \emph{{The Higgs Hunter's
  Guide}}, vol.~80 (2000),
  \href{https://doi.org/10.1201/9780429496448}{10.1201/9780429496448}.

\bibitem{Deshpande:1977rw}
N.G.~Deshpande and E.~Ma, \emph{{Pattern of Symmetry Breaking with Two Higgs
  Doublets}}, \href{https://doi.org/10.1103/PhysRevD.18.2574}{\emph{Phys. Rev.
  D} {\bfseries 18} (1978) 2574}.

\bibitem{Barbieri:2006dq}
R.~Barbieri, L.J.~Hall and V.S.~Rychkov, \emph{{Improved naturalness with a
  heavy Higgs: An Alternative road to LHC physics}},
  \href{https://doi.org/10.1103/PhysRevD.74.015007}{\emph{Phys. Rev. D}
  {\bfseries 74} (2006) 015007}
  [\href{https://arxiv.org/abs/hep-ph/0603188}{{\ttfamily hep-ph/0603188}}].

\bibitem{Keus:2014jha}
V.~Keus, S.F.~King, S.~Moretti and D.~Sokolowska, \emph{{Dark Matter with Two
  Inert Doublets plus One Higgs Doublet}},
  \href{https://doi.org/10.1007/JHEP11(2014)016}{\emph{JHEP} {\bfseries 11}
  (2014) 016} [\href{https://arxiv.org/abs/1407.7859}{{\ttfamily 1407.7859}}].

\bibitem{Keus:2014isa}
V.~Keus, S.F.~King and S.~Moretti, \emph{{Phenomenology of the inert ( 2+1 )
  and ( 4+2 ) Higgs doublet models}},
  \href{https://doi.org/10.1103/PhysRevD.90.075015}{\emph{Phys. Rev. D}
  {\bfseries 90} (2014) 075015}
  [\href{https://arxiv.org/abs/1408.0796}{{\ttfamily 1408.0796}}].

\bibitem{Keus:2015xya}
V.~Keus, S.F.~King, S.~Moretti and D.~Sokolowska, \emph{{Observable Heavy Higgs
  Dark Matter}}, \href{https://doi.org/10.1007/JHEP11(2015)003}{\emph{JHEP}
  {\bfseries 11} (2015) 003}
  [\href{https://arxiv.org/abs/1507.08433}{{\ttfamily 1507.08433}}].

\bibitem{Cordero-Cid:2016krd}
A.~Cordero-Cid, J.~Hern\'andez-S\'anchez, V.~Keus, S.F.~King, S.~Moretti,
  D.~Rojas et~al., \emph{{CP violating scalar Dark Matter}},
  \href{https://doi.org/10.1007/JHEP12(2016)014}{\emph{JHEP} {\bfseries 12}
  (2016) 014} [\href{https://arxiv.org/abs/1608.01673}{{\ttfamily
  1608.01673}}].

\bibitem{Cordero-Cid:2018man}
A.~Cordero-Cid, J.~Hern\'andez-S\'anchez, V.~Keus, S.~Moretti, D.~Rojas and
  D.~Soko\l{}owska, \emph{{Lepton collider indirect signatures of dark
  CP-violation}},
  \href{https://doi.org/10.1140/epjc/s10052-020-7689-0}{\emph{Eur. Phys. J. C}
  {\bfseries 80} (2020) 135}
  [\href{https://arxiv.org/abs/1812.00820}{{\ttfamily 1812.00820}}].

\bibitem{Cordero-Cid:2020yba}
A.~Cordero-Cid, J.~Hern\'andez-S\'anchez, V.~Keus, S.~Moretti, D.~Rojas-Ciofalo
  and D.~Soko\l{}owska, \emph{{Collider signatures of dark $CP$-violation}},
  \href{https://doi.org/10.1103/PhysRevD.101.095023}{\emph{Phys. Rev. D}
  {\bfseries 101} (2020) 095023}
  [\href{https://arxiv.org/abs/2002.04616}{{\ttfamily 2002.04616}}].

\bibitem{Hernandez-Sanchez:2020aop}
J.~Hernandez-Sanchez, V.~Keus, S.~Moretti, D.~Rojas-Ciofalo and D.~Sokolowska,
  \emph{{Complementary Probes of Two-component Dark Matter}},
  \href{https://arxiv.org/abs/2012.11621}{{\ttfamily 2012.11621}}.

\bibitem{Hernandez-Otero:2022dxd}
D.~Hern\'andez-Otero, J.~Hern\'andez-S\'anchez, S.~Moretti and T.~Shindou,
  \emph{{The $Z_3$ soft breaking in the I(2+1)HDM and its probes at present and
  future colliders}},  \href{https://arxiv.org/abs/2203.06323}{{\ttfamily
  2203.06323}}.

\bibitem{Dey:2023exa}
A.~Dey, V.~Keus, S.~Moretti and C.~Shepherd-Themistocleous, \emph{{A smoking
  gun signature of the 3HDM}},
  \href{https://doi.org/10.1007/JHEP07(2024)038}{\emph{JHEP} {\bfseries 07}
  (2024) 038} [\href{https://arxiv.org/abs/2310.06593}{{\ttfamily
  2310.06593}}].

\bibitem{Keus:2013hya}
V.~Keus, S.F.~King and S.~Moretti, \emph{{Three-Higgs-doublet models:
  symmetries, potentials and Higgs boson masses}},
  \href{https://doi.org/10.1007/JHEP01(2014)052}{\emph{JHEP} {\bfseries 01}
  (2014) 052} [\href{https://arxiv.org/abs/1310.8253}{{\ttfamily 1310.8253}}].

\bibitem{Moretti:2015cwa}
S.~Moretti and K.~Yagyu, \emph{{Constraints on Parameter Space from
  Perturbative Unitarity in Models with Three Scalar Doublets}},
  \href{https://doi.org/10.1103/PhysRevD.91.055022}{\emph{Phys. Rev. D}
  {\bfseries 91} (2015) 055022}
  [\href{https://arxiv.org/abs/1501.06544}{{\ttfamily 1501.06544}}].

\bibitem{Merchand:2019bod}
M.~Merchand and M.~Sher, \emph{{Constraints on the Parameter Space in an Inert
  Doublet Model with two Active Doublets}},
  \href{https://doi.org/10.1007/JHEP03(2020)108}{\emph{JHEP} {\bfseries 03}
  (2020) 108} [\href{https://arxiv.org/abs/1911.06477}{{\ttfamily
  1911.06477}}].

\bibitem{Khater:2021wcx}
W.~Khater, A.~Kun\v{c}inas, O.M.~Ogreid, P.~Osland and M.N.~Rebelo, \emph{{Dark
  matter in three-Higgs-doublet models with S$_{3}$ symmetry}},
  \href{https://doi.org/10.1007/JHEP01(2022)120}{\emph{JHEP} {\bfseries 01}
  (2022) 120} [\href{https://arxiv.org/abs/2108.07026}{{\ttfamily
  2108.07026}}].

\bibitem{Kuncinas:2023pub}
A.~Kuncinas, \emph{{Status of dark matter in $S_3$-symmetric 3HDM}},
  \href{https://doi.org/10.22323/1.436.0032}{\emph{PoS} {\bfseries CORFU2022}
  (2023) 032}.

\bibitem{Bento:2017eti}
M.P.~Bento, H.E.~Haber, J.C.~Rom{\~a}o and J.P.~Silva, \emph{{Multi-Higgs
  doublet models: physical parametrization, sum rules and unitarity bounds}},
  \href{https://doi.org/10.1007/JHEP11(2017)095}{\emph{JHEP} {\bfseries 11}
  (2017) 095} [\href{https://arxiv.org/abs/1708.09408}{{\ttfamily
  1708.09408}}].

\bibitem{Chakraborti:2021bpy}
M.~Chakraborti, D.~Das, M.~Levy, S.~Mukherjee and I.~Saha, \emph{{Prospects for
  light charged scalars in a three-Higgs-doublet model with Z3 symmetry}},
  \href{https://doi.org/10.1103/PhysRevD.104.075033}{\emph{Phys. Rev. D}
  {\bfseries 104} (2021) 075033}
  [\href{https://arxiv.org/abs/2104.08146}{{\ttfamily 2104.08146}}].

\bibitem{Aranda:2019vda}
A.~Aranda, D.~Hern\'andez-Otero, J.~Hern\'andez-Sanchez, V.~Keus, S.~Moretti,
  D.~Rojas-Ciofalo et~al., \emph{{Z$_3$ symmetric inert ( 2+1 )-Higgs-doublet
  model}}, \href{https://doi.org/10.1103/PhysRevD.103.015023}{\emph{Phys. Rev.
  D} {\bfseries 103} (2021) 015023}
  [\href{https://arxiv.org/abs/1907.12470}{{\ttfamily 1907.12470}}].

\bibitem{Cabrera:2019gaq}
M.E.~Cabrera, J.A.~Casas, A.~Delgado and S.~Robles, \emph{{Generalized Blind
  Spots for Dark Matter Direct Detection in the 2HDM}},
  \href{https://doi.org/10.1007/JHEP02(2020)166}{\emph{JHEP} {\bfseries 02}
  (2020) 166} [\href{https://arxiv.org/abs/1912.01758}{{\ttfamily
  1912.01758}}].

\bibitem{Arcadi:2025sxc}
G.~Arcadi and S.~Profumo, \emph{{WIMPs Below the Radar: Blind Spots and
  Benchmarks Beyond the Neutrino Floor}},
  \href{https://arxiv.org/abs/2506.19062}{{\ttfamily 2506.19062}}.

\bibitem{Glashow:1976nt}
S.L.~Glashow and S.~Weinberg, \emph{{Natural Conservation Laws for Neutral
  Currents}}, \href{https://doi.org/10.1103/PhysRevD.15.1958}{\emph{Phys. Rev.
  D} {\bfseries 15} (1977) 1958}.

\bibitem{Akeroyd:2016ssd}
A.G.~Akeroyd, S.~Moretti, K.~Yagyu and E.~Yildirim, \emph{{Light charged Higgs
  boson scenario in 3-Higgs doublet models}},
  \href{https://doi.org/10.1142/S0217751X17501457}{\emph{Int. J. Mod. Phys. A}
  {\bfseries 32} (2017) 1750145}
  [\href{https://arxiv.org/abs/1605.05881}{{\ttfamily 1605.05881}}].

\bibitem{Kannike:2012pe}
K.~Kannike, \emph{{Vacuum Stability Conditions From Copositivity Criteria}},
  \href{https://doi.org/10.1140/epjc/s10052-012-2093-z}{\emph{Eur. Phys. J. C}
  {\bfseries 72} (2012) 2093}
  [\href{https://arxiv.org/abs/1205.3781}{{\ttfamily 1205.3781}}].

\bibitem{Chakrabortty:2013mha}
J.~Chakrabortty, P.~Konar and T.~Mondal, \emph{{Copositive Criteria and
  Boundedness of the Scalar Potential}},
  \href{https://doi.org/10.1103/PhysRevD.89.095008}{\emph{Phys. Rev. D}
  {\bfseries 89} (2014) 095008}
  [\href{https://arxiv.org/abs/1311.5666}{{\ttfamily 1311.5666}}].

\bibitem{Das:2019yad}
D.~Das and I.~Saha, \emph{{Alignment limit in three Higgs-doublet models}},
  \href{https://doi.org/10.1103/PhysRevD.100.035021}{\emph{Phys. Rev. D}
  {\bfseries 100} (2019) 035021}
  [\href{https://arxiv.org/abs/1904.03970}{{\ttfamily 1904.03970}}].

\bibitem{Peskin:1991sw}
M.E.~Peskin and T.~Takeuchi, \emph{{Estimation of oblique electroweak
  corrections}}, \href{https://doi.org/10.1103/PhysRevD.46.381}{\emph{Phys.
  Rev. D} {\bfseries 46} (1992) 381}.

\bibitem{Grimus:2008nb}
W.~Grimus, L.~Lavoura, O.M.~Ogreid and P.~Osland, \emph{{The Oblique parameters
  in multi-Higgs-doublet models}},
  \href{https://doi.org/10.1016/j.nuclphysb.2008.04.019}{\emph{Nucl. Phys. B}
  {\bfseries 801} (2008) 81} [\href{https://arxiv.org/abs/0802.4353}{{\ttfamily
  0802.4353}}].

\bibitem{Grimus:2007if}
W.~Grimus, L.~Lavoura, O.M.~Ogreid and P.~Osland, \emph{{A Precision constraint
  on multi-Higgs-doublet models}},
  \href{https://doi.org/10.1088/0954-3899/35/7/075001}{\emph{J. Phys. G}
  {\bfseries 35} (2008) 075001}
  [\href{https://arxiv.org/abs/0711.4022}{{\ttfamily 0711.4022}}].

\bibitem{Bertolini:1985ia}
S.~Bertolini, \emph{{Quantum Effects in a Two Higgs Doublet Model of the
  Electroweak Interactions}},
  \href{https://doi.org/10.1016/0550-3213(86)90341-X}{\emph{Nucl. Phys. B}
  {\bfseries 272} (1986) 77}.

\bibitem{ParticleDataGroup:2022pth}
{\scshape Particle Data Group} collaboration, \emph{{Review of Particle
  Physics}}, \href{https://doi.org/10.1093/ptep/ptac097}{\emph{PTEP} {\bfseries
  2022} (2022) 083C01}.

\bibitem{Batra:2025amk}
N.~Batra, B.~Coleppa, A.~Khanna, S.K.~Rai and A.~Sarkar, \emph{{Constraining
  the 3HDM parameter space using active learning}},
  \href{https://doi.org/10.1103/t5df-67wh}{\emph{Phys. Rev. D} {\bfseries 112}
  (2025) 015011} [\href{https://arxiv.org/abs/2504.07489}{{\ttfamily
  2504.07489}}].

\bibitem{Coleppa:2025qst}
B.~Coleppa, A.~Khanna and G.B.~Krishna, \emph{{3HDM at the ILC}},
  \href{https://arxiv.org/abs/2506.24094}{{\ttfamily 2506.24094}}.

\bibitem{Hmissou:2025uep}
A.~Hmissou, S.~Moretti and L.~Rahili, \emph{{Investigating the 95 GeV Higgs
  Boson Excesses within the I(1+2)HDM}},
  \href{https://arxiv.org/abs/2502.03631}{{\ttfamily 2502.03631}}.

\bibitem{TevatronElectroweakWorkingGroup:2010mao}
{\scshape Tevatron Electroweak Working Group} collaboration, \emph{{Combination
  of CDF and D0 Results on the Width of the W boson}},
  \href{https://arxiv.org/abs/1003.2826}{{\ttfamily 1003.2826}}.

\bibitem{Boline:2011qf}
{\scshape D0} collaboration, \emph{{Measurement of the $W$ Boson Mass and Width
  at the D0 Experiment}},  in \emph{{Meeting of the APS Division of Particles
  and Fields}}, 10, 2011 [\href{https://arxiv.org/abs/1110.1093}{{\ttfamily
  1110.1093}}].

\bibitem{ALEPH:2013dgf}
{\scshape ALEPH, DELPHI, L3, OPAL, LEP Electroweak} collaboration,
  \emph{{Electroweak Measurements in Electron-Positron Collisions at
  W-Boson-Pair Energies at LEP}},
  \href{https://doi.org/10.1016/j.physrep.2013.07.004}{\emph{Phys. Rept.}
  {\bfseries 532} (2013) 119}
  [\href{https://arxiv.org/abs/1302.3415}{{\ttfamily 1302.3415}}].

\bibitem{Lundstrom:2008ai}
E.~Lundstrom, M.~Gustafsson and J.~Edsjo, \emph{{The Inert Doublet Model and
  LEP II Limits}},
  \href{https://doi.org/10.1103/PhysRevD.79.035013}{\emph{Phys. Rev. D}
  {\bfseries 79} (2009) 035013}
  [\href{https://arxiv.org/abs/0810.3924}{{\ttfamily 0810.3924}}].

\bibitem{ALEPH:2013htx}
{\scshape ALEPH, DELPHI, L3, OPAL, LEP} collaboration, \emph{{Search for
  Charged Higgs bosons: Combined Results Using LEP Data}},
  \href{https://doi.org/10.1140/epjc/s10052-013-2463-1}{\emph{Eur. Phys. J. C}
  {\bfseries 73} (2013) 2463}
  [\href{https://arxiv.org/abs/1301.6065}{{\ttfamily 1301.6065}}].

\bibitem{HFLAV:2016hnz}
{\scshape HFLAV} collaboration, \emph{{Averages of $b$-hadron, $c$-hadron, and
  $\tau$-lepton properties as of summer 2016}},
  \href{https://doi.org/10.1140/epjc/s10052-017-5058-4}{\emph{Eur. Phys. J. C}
  {\bfseries 77} (2017) 895}
  [\href{https://arxiv.org/abs/1612.07233}{{\ttfamily 1612.07233}}].

\bibitem{Misiak:2017bgg}
M.~Misiak and M.~Steinhauser, \emph{{Weak radiative decays of the B meson and
  bounds on $M_{H^\pm }$ in the Two-Higgs-Doublet Model}},
  \href{https://doi.org/10.1140/epjc/s10052-017-4776-y}{\emph{Eur. Phys. J. C}
  {\bfseries 77} (2017) 201}
  [\href{https://arxiv.org/abs/1702.04571}{{\ttfamily 1702.04571}}].

\bibitem{Arbey:2017gmh}
A.~Arbey, F.~Mahmoudi, O.~Stal and T.~Stefaniak, \emph{{Status of the Charged
  Higgs Boson in Two Higgs Doublet Models}},
  \href{https://doi.org/10.1140/epjc/s10052-018-5651-1}{\emph{Eur. Phys. J. C}
  {\bfseries 78} (2018) 182}
  [\href{https://arxiv.org/abs/1706.07414}{{\ttfamily 1706.07414}}].

\bibitem{Alloul:2013bka}
A.~Alloul, N.D.~Christensen, C.~Degrande, C.~Duhr and B.~Fuks, \emph{{FeynRules
  2.0 - A complete toolbox for tree-level phenomenology}},
  \href{https://doi.org/10.1016/j.cpc.2014.04.012}{\emph{Comput. Phys. Commun.}
  {\bfseries 185} (2014) 2250}
  [\href{https://arxiv.org/abs/1310.1921}{{\ttfamily 1310.1921}}].

\bibitem{Belanger:2013oya}
G.~Belanger, F.~Boudjema, A.~Pukhov and A.~Semenov, \emph{{micrOMEGAs$\_$3: A
  program for calculating dark matter observables}},
  \href{https://doi.org/10.1016/j.cpc.2013.10.016}{\emph{Comput. Phys. Commun.}
  {\bfseries 185} (2014) 960}
  [\href{https://arxiv.org/abs/1305.0237}{{\ttfamily 1305.0237}}].

\bibitem{LZ:2024zvo}
{\scshape LZ} collaboration, \emph{{Dark Matter Search Results from
  4.2{\,}{\,}Tonne-Years of Exposure of the LUX-ZEPLIN (LZ) Experiment}},
  \href{https://doi.org/10.1103/4dyc-z8zf}{\emph{Phys. Rev. Lett.} {\bfseries
  135} (2025) 011802} [\href{https://arxiv.org/abs/2410.17036}{{\ttfamily
  2410.17036}}].

\bibitem{He:2008qm}
X.-G.~He, T.~Li, X.-Q.~Li, J.~Tandean and H.-C.~Tsai, \emph{{Constraints on
  Scalar Dark Matter from Direct Experimental Searches}},
  \href{https://doi.org/10.1103/PhysRevD.79.023521}{\emph{Phys. Rev. D}
  {\bfseries 79} (2009) 023521}
  [\href{https://arxiv.org/abs/0811.0658}{{\ttfamily 0811.0658}}].

\bibitem{Storn:1997uea}
R.~Storn and K.~Price, \emph{{Differential Evolution \textendash{} A Simple and
  Efficient Heuristic for global Optimization over Continuous Spaces}},
  \href{https://doi.org/10.1023/A:1008202821328}{\emph{J. Global Optim.}
  {\bfseries 11} (1997) 341}.

\bibitem{AbdusSalam:2020rdj}
S.S.~AbdusSalam et~al., \emph{{Simple and statistically sound recommendations
  for analysing physical theories}},
  \href{https://doi.org/10.1088/1361-6633/ac60ac}{\emph{Rept. Prog. Phys.}
  {\bfseries 85} (2022) 052201}
  [\href{https://arxiv.org/abs/2012.09874}{{\ttfamily 2012.09874}}].

\bibitem{Ellis:2017ndg}
J.~Ellis, A.~Fowlie, L.~Marzola and M.~Raidal, \emph{{Statistical Analyses of
  Higgs- and Z-Portal Dark Matter Models}},
  \href{https://doi.org/10.1103/PhysRevD.97.115014}{\emph{Phys. Rev. D}
  {\bfseries 97} (2018) 115014}
  [\href{https://arxiv.org/abs/1711.09912}{{\ttfamily 1711.09912}}].

\bibitem{Yaguna:2024jor}
C.E.~Yaguna and O.~Zapata, \emph{{Singlet Dirac dark matter streamlined}},
  \href{https://doi.org/10.1088/1475-7516/2024/06/049}{\emph{JCAP} {\bfseries
  06} (2024) 049} [\href{https://arxiv.org/abs/2401.13101}{{\ttfamily
  2401.13101}}].

\bibitem{Wilks:1938dza}
S.S.~Wilks, \emph{{The Large-Sample Distribution of the Likelihood Ratio for
  Testing Composite Hypotheses}},
  \href{https://doi.org/10.1214/aoms/1177732360}{\emph{Annals Math. Statist.}
  {\bfseries 9} (1938) 60}.

\bibitem{LZ:2022lsv}
{\scshape LZ} collaboration, \emph{{First Dark Matter Search Results from the
  LUX-ZEPLIN (LZ) Experiment}},
  \href{https://doi.org/10.1103/PhysRevLett.131.041002}{\emph{Phys. Rev. Lett.}
  {\bfseries 131} (2023) 041002}
  [\href{https://arxiv.org/abs/2207.03764}{{\ttfamily 2207.03764}}].

\end{thebibliography}\endgroup
\end{document}